\documentclass[fleqn,usenatbib]{mnras}

\usepackage{newtxtext,newtxmath}

\usepackage[T1]{fontenc}

\DeclareRobustCommand{\VAN}[3]{#2}
\let\VANthebibliography\thebibliography
\def\thebibliography{\DeclareRobustCommand{\VAN}[3]{##3}\VANthebibliography}

\usepackage{graphicx}	
\usepackage{amsmath}	
\usepackage{pdflscape}
\usepackage{afterpage}
\usepackage{orcidlink}

\newcommand{\chandra}{\textit{Chandra}}
\newcommand{\nustar}{\textit{NuSTAR}}
\newcommand{\suzaku}{{\it Suzaku}}

\newcommand{\athena}{{\it Athena}}
\newcommand{\xmm}{\textit{XMM-Newton}}

\newcommand{\xrism}{{\it XRISM}}

\newcommand{\fv}{f_{\rm v}}
\newcommand{\lx}{L_{\rm X}}

\newcommand{\mout}{\dot M_{\rm out}}

\newcommand{\ledd}{L_{\rm Edd}}
\newcommand{\medd}{\dot M_{\rm Edd}}

\title[The X-ray Disk/Wind Degeneracy in AGN]{The X-ray Disk/Wind Degeneracy in AGN}

\author[M. L. Parker et al.]{M. L. Parker,$^{1}$\thanks{E-mail: mlparker@ast.cam.ac.uk}\orcidlink{0000-0002-8466-7317}
G. A. Matzeu,$^{2,3}$\orcidlink{0000-0003-1994-5322}
J. H. Matthews,$^1$\orcidlink{0000-0002-3493-7737}
M. J. Middleton,$^4$\orcidlink{0000-0002-8183-2970}
T. Dauser,$^5$\orcidlink{0000-0003-4583-9048}\newauthor
J. Jiang$^6$\orcidlink{0000-0002-9639-4352} and
A. M. Joyce.$^5$\orcidlink{0000-0001-5437-8541}
\\
$^{1}$Institute of Astronomy, University of Cambridge, Madingley Road, Cambridge, CB3 0HA, UK\\
$^2$Department of Physics and Astronomy (DIFA), University of Bologna, Via Gobetti, 93/2, I-40129 Bologna, Italy\\
$^{3}$INAF-Osservatorio di Astrofisica e Scienza dello Spazio di Bologna, Via Gobetti, 93/3, I-40129 Bologna, Italy\\
$^4$Department of Physics and Astronomy, University of Southampton, Southampton, SO17 1BJ, UK\\
$^{5}$Dr. Karl Remeis-Observatory \& ECAP, University of Erlangen-Nuremberg,    Sternwartstr. 7, 96049 Bamberg, Germany\\
$^6$Department of Astronomy, Tsinghua Univerisity, Shuangqing Road, Beĳing 100084, China\\
}

\date{Accepted XXX. Received YYY; in original form ZZZ}

\pubyear{2021}

\begin{document}
\label{firstpage}
\pagerange{\pageref{firstpage}--\pageref{lastpage}}
\maketitle
\begin{abstract}
Relativistic Fe~K emission lines from accretion disks and from disk winds encode key information about black holes, and their accretion and feedback mechanisms. We show that these two processes can in principle produce indistinguishable line profiles, such that they cannot be disentangled spectrally. We argue that it is likely that in many cases both processes contribute to the net line profile, and their relative contributions cannot be constrained purely by Fe~K spectroscopy. 
In almost all studies of Fe~K emission to date, a single process (either disk reflection or wind Compton scattering) is assumed to dominate the total line profile. We demonstrate that fitting a single process emission model (pure reflection or pure wind) to a hybrid line profile results in large systematic biases in the estimates of key parameters, such as mass outflow rate and spin. We discuss various strategies to mitigate this effect, such as including high energy data covering the Compton hump, and the implications for future X-ray missions.
\end{abstract}



\section{Introduction}
The relativistic iron line is a key feature of active galactic nuclei (AGN) X-ray spectra, observationally characterised as an emission line between 6.4--7~keV, with a very high degree of blue- and redshift, requiring relativistic effects and/or velocities. In most cases, these lines are interpreted as emission from the X-ray corona that is reprocessed off the surface of the inner accretion disk, known as the reflection spectrum. This produces a characteristic spectrum of fluorescent emission lines, most notably the Fe~K line, which are blurred and skewed by the extreme velocity of the inner accretion disk and the gravitional redshift of the black hole potential well \citep[e.g.][]{Fabian89,George91,Tanaka95}. This relativistic blurring encodes a great deal of information about the black hole system, in particular the spin of the black hole, $a$. At high spin, the innermost stable circular orbit (ISCO) approaches closer to the event horizon. This means that the disk reflection spectrum can contain more strongly redshifted emission from deeper in the potential well. In principle, it is therefore possible to infer the spin by measuring the degree of redshift in the reflection spectrum, in particular in the profile of the relativistic Fe~K line. This parameter is of great interest to the wider community, as it tells us about the formation and growth channels of black holes \citep[see e.g. review by][]{Reynolds21}, and measuring spin through this method is a common goal of future X-ray missions like \xrism\ and \athena .

With the launch of \chandra\ and \xmm , a second type of relativistic phenomenon was identified in AGN X-ray spectra: the so-called ultra-fast outflows (UFOs). The signatures of these UFOs are strongly blueshifted absorption lines of highly ionised species, most commonly Fe~\textsc{xxv--xxvi} \citep[][]{Chartas02,Pounds03, Reeves03}. These absorption lines are generally interpreted as evidence for a wind, launched from the surface of the accretion disk and accelerated by magnetic fields and/or radiation to moderately relativistic velocities (0.1--0.3$c$). With their extreme velocities and the implied high kinetic power and momentum, UFOs are excellent candidates for driving AGN feedback \citep[see e.g. review by][]{Fabian12_feedback}.

However, as they are frequently detected through an absorption component only provide information about the line of sight absorption and this restricted sight line makes it extremely difficult to determine the global properties of such a disk wind. As the emission gives a quantity which is integrated over some volume it can, in principle, provide information about the wind geometry (essentially, the absorption encodes information about the sightline that intercepts the wind, while the emission is averaged from all angles). The emission mechanism is similar to the reflection case: X-rays from the corona, close to the black hole, hit the wind and are reprocessed into photoionised emission lines, and either red- or blue-shifted according to the velocity of the wind surface relative to the line of sight. In certain regions of parameter space, this scenario predicts a P-Cygni like profile \citep[e.g.][]{Done07, Sim08, Sim10, Nardini15}, where the combined emission and absorption profile encodes information like the covering factor, mass outflow rate, the ionisation structure, and the acceleration profile of the wind.

A clear problem arises here. Both of these methods promise groundbreaking results with great importance for astronomy, and both strongly rely on inferring properties of their respective physical process from the shape of the Fe~K emission line profile. Both techniques have been used with great success, but in almost every case they do so assuming that only one process contributes to the total Fe~K emission line profile. It is immediately obvious that the presence of un-modelled emission from either disk reflection or a disk wind will introduce a large systematic error into parameters inferred by fitting the line profile assuming that only the other process contributes. As the detection of a disk wind clearly implies the existence of an accretion disk and its associated emission, it is entirely possible that in many cases both processes do contribute simultaneously and that the degeneracy between them is the dominant source of systematic error for both fields.

In this paper, we attempt to estimate the scale of the systematic error introduced by this degeneracy by considering fits to archetypal sources, and through simulations. The paper is organised as follows:
\begin{itemize}
    \item In Section~\ref{sec:archetypalsources} we consider fits to real data from AGN that are prototypical wind or reflection sources, showing that they can easily be fit with the alternative model.
    \item In Section~\ref{sec:simulations} we set up various simulations to explore the degeneracy between the reflection and wind models, and the effect it has on parameter estimation. In~\ref{sec:controls} we consider a set of control simulations, to see how well parameters are recovered when the reflection model is used to fit simulated reflection spectra, and a disk wind model is used to fit disk wind spectra. We then go on to consider a hybrid model in~\ref{sec:hybridsims}, where both reflection and a wind contribute. We examine the parameter recovery when either the reflection or disk wind model is used to fit these hybrid spectra.
    \item In Section~\ref{sec:discussion} we discuss the implications of these results for \xrism\ and \athena\ (\ref{sec:discuss_xrism}), the impact on disk wind and reflection spectroscopy (\ref{sec:discuss_diskwind} and \ref{sec:discuss_reflection}), strategies for mitigating this effect (\ref{sec:discuss_mitigation}), and caveats to the analysis (\ref{sec:discuss_caveats}).
    \item Finally, in Section~\ref{sec:conclusions} we present our conclusions.
\end{itemize}

\section{Fits to archetypal sources}
\label{sec:archetypalsources}
To demonstrate the scale of the problem, we consider two sources that have been consistently modelled with one of the two models, with relatively little consideration of the other: PDS~456 and IRAS~13224-3809. We do not argue that the prevailing interpretation is necessarily incorrect in either case, instead we argue that it is extremely difficult, if not impossible, to accurately constrain the relative contributions of disk reflection and wind scattering to the net iron line profile from spectroscopy alone.

\subsection{PDS~456}

\begin{figure*}
    \centering
    \includegraphics[width=14cm]{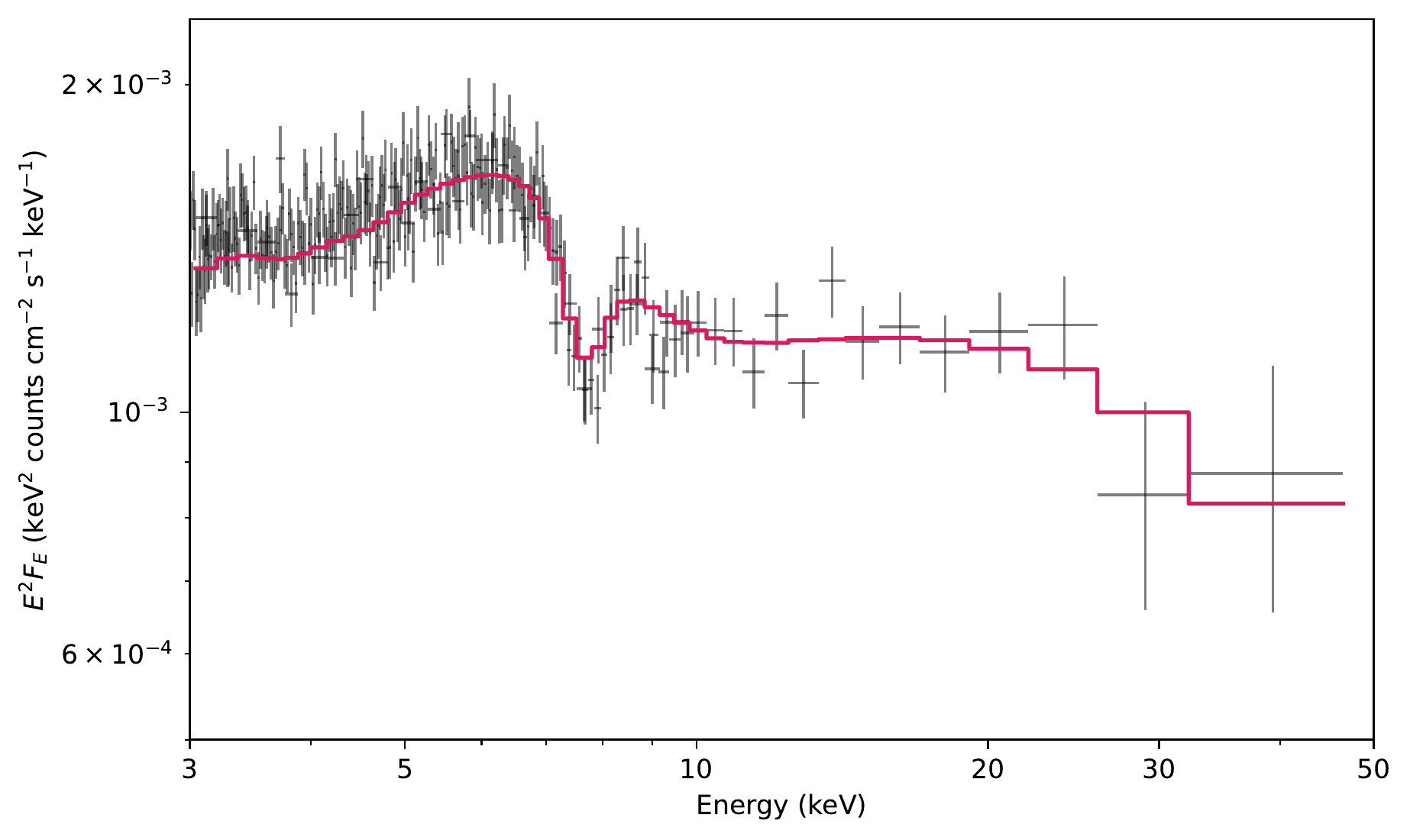}
    \caption{Left: 3--50 keV (observed frame) spectrum of PDS~456 with \xmm\ and \nustar , fit with a reflection only model and a single Gaussian absorption line. A good fit to the iron emission line is obtained with no need to account for wind emission.}
    \label{fig:pds}
\end{figure*}

PDS 456 is a low redshift ($z = 0.184$) radio quiet quasar, which hosts one of the earliest detected UFOs \citep[][]{Reeves03} and has clear evidence for multiple layers of absorption \citep[][]{Reeves16, Reeves18_pds456, Haerer21}. As well as the absorption, a strong, broad Fe emission line is present from $\sim6$--8 keV (rest frame). \citet{Nardini15} model this as a P-Cygni profile, where the Fe~K emission is produced by the wind, and argue that the contribution from disk reflection must be weak as a strong Compton hump is not visible in the \nustar\ spectrum. \citet{Chiang17} were able to fit the archival \suzaku\ and \nustar\ spectra with a pure reflection model plus two layers of ionised absorption, and found a reverberation lag, arguing for the presence of at least some reflection emission. We fit the same \xmm\ and \nustar\ spectra as shown in Fig.~3 of \citeauthor{Nardini15} with a pure reflection model \citep[\textsc{relxill},][]{Garcia14} and a Gaussian absorption line. This corresponds to a scenario where a wind is present and absorbs the emission from the inner disk, but the net emission from the wind is negligible compared to that from the disk (for example, because the solid angle of the wind is small, or most of it is over-ionised). We find an excellent fit to the spectrum (Fig~\ref{fig:pds}, parameters in Table~\ref{tab:pds_fit}), including the high energy band, which is consistent with the presence of a weak Compton hump.

\begin{table}
    \centering
    \caption{Best fit parameters for the pure reflection (with Gaussian absorption line) fit to the high energy spectrum of PDS~456.}
    \label{tab:pds_fit}
    \begin{tabular}{l c r}
    \hline
        Parameter    &  Value   & Description  \\
    \hline
        $E_\mathrm{Gauss}$ & $9.08_{-0.04}^{+0.05}$ & Line energy (keV)\\
        $\sigma_\mathrm{Gauss}$ & $0.29\pm0.05$ & Line width (keV)\\
        $S_\mathrm{Gauss}$ & $0.22\pm0.04$ & Line strength\\
        $q_\mathrm{in}$ & $>3.5$ & Emissivity index\\
        $a$ & $0.97_{-0.03}^{+0.01}$ & Spin\\
        $i$ & $70_{-6}^{+3}$ & Inclination (degrees)\\
        $\Gamma$ & $1.83_{-0.06}^{+0.05}$ & Photon index\\
        $\log(\xi)$ & $3.7_{-0.1}^{+0.3}$ & Ionisation (erg~cm~s$^{-1}$)\\
        $A_\mathrm{Fe}$ & $>4$ & Iron abundance (solar)\\
        $E_\mathrm{cut}$ & $46_{-9}^{+12}$ & High energy cutoff (keV)\\
        $R$ & $1.9_{-0.8}^{+5.8}$ & Reflection fraction\\
        \hline
    \end{tabular}
\end{table}

This does not mean that the iron K line in PDS~456 is necessarily predominantly produced by reflection, but it does mean that the disk wind model previously used to fit the emission is extremely degenerate with the disk reflection model. Since the existence of a disk wind presupposes the existence of a disk, we have a strong prior that at least some of the emission in the line profile must be coming from the disk. Without independent constraints on this, it is impossible to say which emission in the line profile is due to the wind, and therefore impossible to reliably measure any property of the wind from the emission line.


\subsection{IRAS~13224-3809}

IRAS~13224-3809 is a narrow line Seyfert 1 (NLS1) AGN, best known for its remarkable variability and strong Fe~K emission line \citep[e.g.][]{Fabian12}. In particular, IRAS~13224 shows good evidence for a Fe~K lag \citep[][]{Kara13_iras}, and evidence for an Fe~L line in the soft excess \citep[][]{Jiang18}
In 2017, we demonstrated that a strong flux dependent \citep[][]{Parker17_nature} and velocity dependent \citep[][]{Pinto18} UFO absorption line is present in the Fe~K band. 

\begin{figure*}
    \centering
    \includegraphics[width=0.45\linewidth]{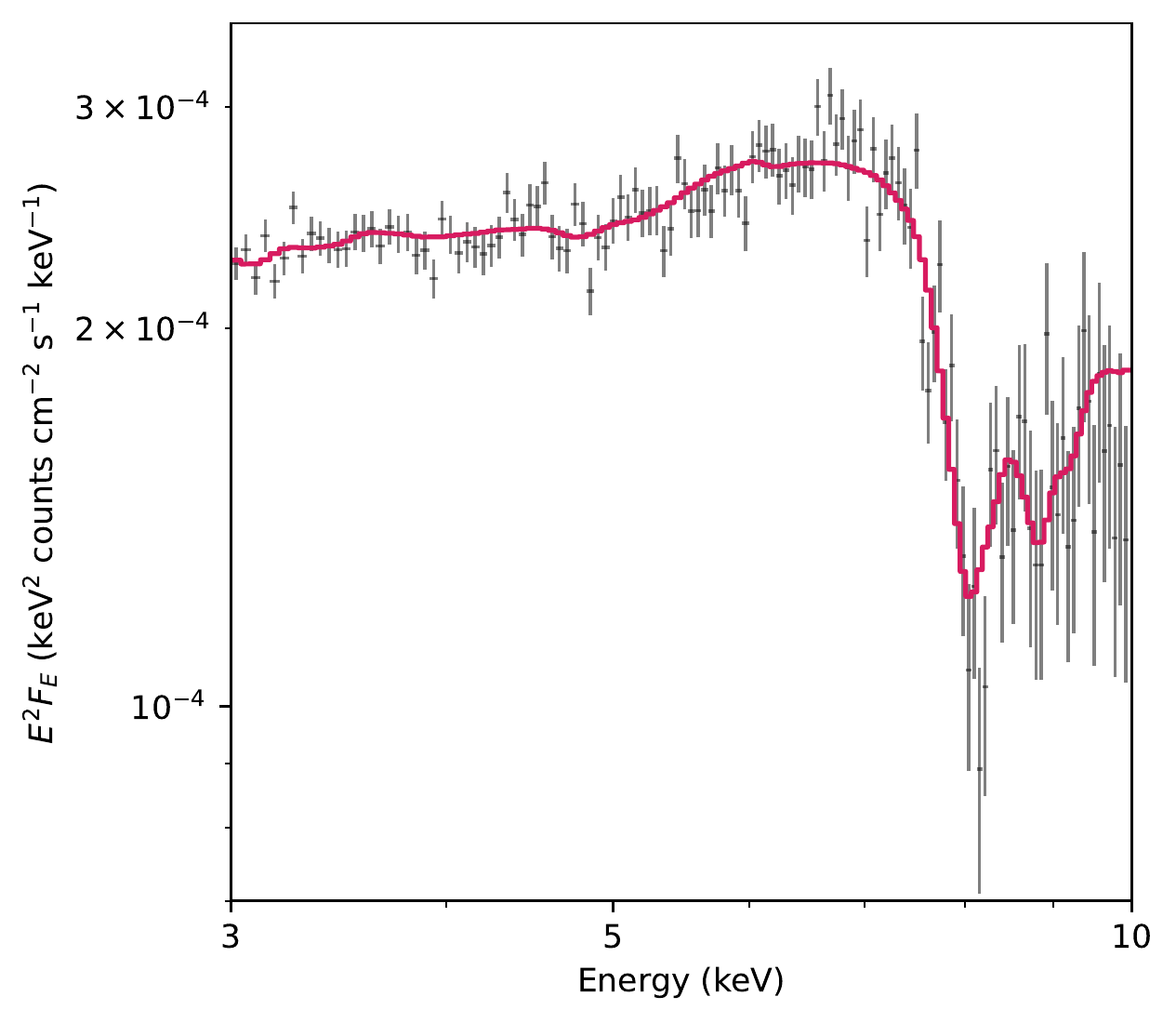}
    \includegraphics[width=0.45\linewidth]{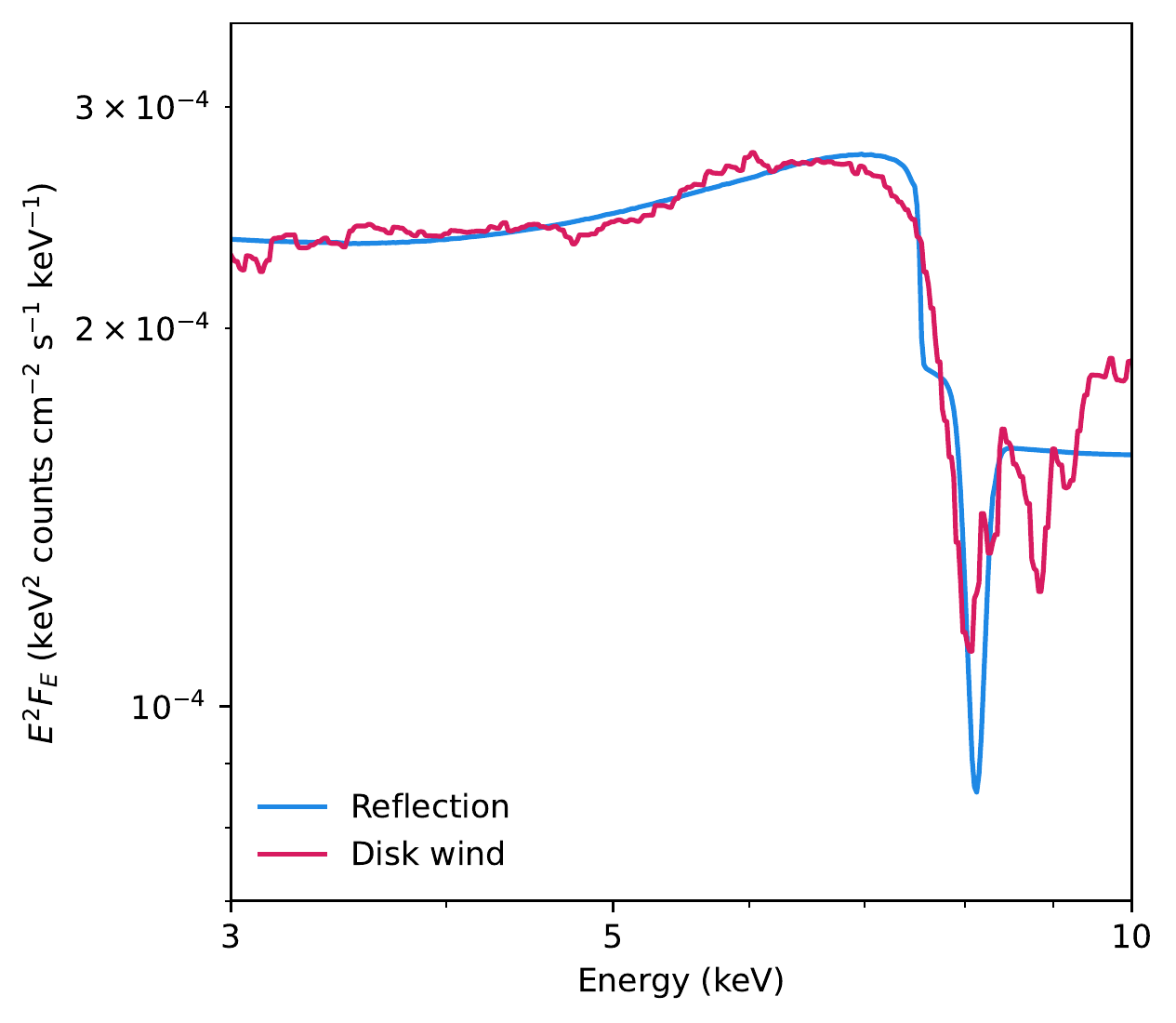}
    \caption{Left: 3--10 keV spectrum of IRAS~13224-3809, fit with a disk wind only model. The spectrum is corrected for the instrumental effective area, but not unfolded from the response. A good fit to the iron emission line is obtained with no need to account for relativistic reflection. Right: A comparison of the 3--10 keV disk wind model to the reflection model with a single Gaussian line from \citet{Parker17_nature}. The shape of the iron line profile is almost identical in the two models.}
    \label{fig:iras}
\end{figure*}

In previous work, we have used a relativistic reflection model to describe the iron line, neglecting emission from the wind and including a separate absorption component to describe the absorption line. However, it is entirely possible that some of the net emission line profile is produced by emission scattered off the wind. To test this, we attempt to model the Fe~K emission and absorption with a pure disk wind model.

In this test we physically model the spectral imprints from the wind with a custom generated table of simulated spectra using a Monte Carlo radiative transfer code including a parameterised disk wind model \citep{Sim08,Sim10}. This multiplicative table modifies a continuum, assumed by the simulation to be a powerlaw, adding both absorption and emission components.
The Monte Carlo radiative transfer method \citep[e.g.,][]{Lucy02,Noebauer2019} is used to compute synthetic spectra from a multi-dimensional (3D), smooth, steady-state and bi-conical flow with an opening angle of 45 degrees \citep[][]{Knigge95} containing both the transmitted and scattered radiation through and off the wind respectively. We assume a maximum launch radius of $48\,\rm r_{\rm g}$ ($1.5\times$ the minimum), an X-ray emission region of $6\,\rm r_{\rm g}$ and an outer boundary of the wind of about $1000\,\rm r_{\rm min}$ and an accretion efficiency of $\eta=0.06$.
 
The disk-wind model provides a self consistent treatment of both the absorption and emission components from a more geometrically `plausible' flow that includes a radially dependent ionization and velocity structure across the wind. We are able to probe the overall mass outflow rate, launching radius and terminal velocity of the wind as well as the LOS viewing angle.

In this work, we generated a disk-wind table based on the prototypical wind in PDS\,456, constituted of $5\times6\times4\times20=2400$ synthetic spectra. Here we assume a minimum launching radius of $32\,R_{\rm g}=4.8\times10^{15}\,\rm cm$ for a black hole mass of $10^{9}\,\rm M_{\odot}$ (see Table\,\ref{tab:dw_values} for details). A more extensive version of this model is in development and will be made public soon (Matzeu et al., in prep).

\begin{table*}
  \begin{tabular}{l|c c c c}
\hline

Parameter        &value range                        &$\Delta$ value   &Steps         &Description           \\

\hline

$R_{\rm min}$          &$32$                               &                &           &Launch radius in gravitational radii ($r_{\rm g}$)            \\

$\Gamma$         &$2.1$                              &               &           & Illuminating power law photon index \\

$\dot{M}$            &$0.1$--$0.5$                       &$0.1$            &$5$          &Fraction of mass outflow rate in Eddington units $(\mout/\medd)$\\

$\fv$            &$0.25$--$1.5$                      &$0.25$            &$6$           &Terminal velocity parameter where $v_{\infty}=\fv\sqrt{2GM_{\rm BH}/R}$   \\

$\lx$            &$0.5$--$2.0$                       &$0.5$             &$4$         &$\%$ of ionizing $2$--$10$\,keV luminosity in Eddington units $(L_{2-10\,\rm keV}/\ledd)$\\

$\mu=\cos(i)$ &$0.025$--$0.975$                   &$0.025$            &$20$          &Cosine of LOS wind inclination wrt the polar $(z)$--axis  

\\

\hline

\multicolumn{3}{c}{Total spectra} &2400&\\

\hline

  \end{tabular}
  \caption{Customised disk-wind model parameters based on PDS\,456}
\label{tab:dw_values}
\end{table*}

We fit the 3--10 keV spectrum of IRAS~13224-3809 with a two-zone disk wind model (\textsc{diskwind $\times$ diskwind $\times$ powerlaw}) which physically corresponds to the line of sight intercepting two distinct wind streamlines (a single-zone does not give an acceptable fit, leaving residuals around 9~keV), shown in Fig.~\ref{fig:iras}. The parameters of the zones are fixed, aside from their velocity parameters which are allowed to vary independently. The emission and absorption line profiles are well fit with the two layer model, leaving no systematic residuals. The best-fit parameters are given in Table~\ref{tab:iras_fit}. In the right panel of Fig.~\ref{fig:iras} we show a comparison of the two-zone disk wind model with the reflection plus Gaussian model from \citet{Parker17_nature}. Through the Fe~K profile the two model are almost identical, down to the level of the numerical noise in the disk wind spectrum. Some differences appear at higher energies, where the absorption prescriptions differ, but these are unlikely to offer strong constraints on the model and in practise could easily be accounted for by minor alterations to either model setup.

As with the case of PDS~456, this does not necessarily mean that the prevailing interpretation of the spectrum of IRAS~13224-3809 is incorrect, and it is likely that the inclusion of broad-band data can rule out this pure wind model (see discussion of the Compton hump in \ref{sec:broadband}). However, this test demonstrates that it is extremely difficult to distinguish disk emission from relativistic wind emission in the iron line, and we cannot rule out some contribution to the overall line profile from wind emission in this and similar sources.

\begin{table}
    \centering
    \caption{Best-fit parameters for the pure disk wind fit to the IRAS~13224-3809 Fe~K line profile shown in Fig.~\ref{fig:iras}.}
    \label{tab:iras_fit}
    \begin{tabular}{l c r}
    \hline
        Parameter &  Value   &   Description  \\
    \hline
        $\Gamma$ &  $2.04_{-0.02}^{+0.01}$ & Photon index$^a$\\
        $\dot{M}$ & $>0.5$ & Mass outflow rate in Eddington units\\
        $\mu$ & $0.5$ & Inclination$^b$\\
        $L_\mathrm{X}$ & $1.0\pm0.2$ & Luminosity in Eddington units\\
        $f_\mathrm{v,1}$ & $1.75$ & Velocity parameter$^b$\\
        $f_\mathrm{v,2}$ & $1.25$ & Velocity parameter$^b$\\
    \hline
        
    \end{tabular}

$^a$The index of the illuminating powerlaw in the disk wind model is fixed at 2.1, but we fit the powerlaw slope freely in the model. The value returned by the fit is close enough to the assumed value that it should not introduce significant systematic error.
    
$^b$The interpolation issues between $\mu$ and $f_v$ parameters can be problematic. We initially fit these parameters freely, then tie them to the nearest gridpoint. For this reason, no errors are given on these values as they cannot be reasonably estimated.
\end{table}


\section{Simulations}
\label{sec:simulations}

To investigate the potential impact of degeneracies between disk winds and reflection we consider the error introduced by fitting an iron line spectrum that includes both processes with a model that only accounts for one. This single model fitting approach is standard in X-ray astronomy, and is likely to fail if the underlying spectrum is more complex. In each case, we are looking to see how well the parameters used to simulate the spectrum are recovered by fitting the iron line band (i.e. from 3--10 keV). We note that in principle both models can be constrained by spectral features outside this energy range, however these features (primarily the soft excess and Compton hump) do not offer the same diagnostic power as the iron line, and are easily confused with emission from other processes. We discuss this further in Section~\ref{sec:discuss_mitigation}.

\subsection{Controls}
\label{sec:controls}
To establish the effect of the degeneracy, we first need to know how accurate and precise the parameter recovery is without this effect. We therefore perform a set of control simulations, where a single process contributes, and the same single process model is used to fit the data.

\subsubsection{Reflection}
\label{sec:refcontrol}

Using \textsc{Xspec} \citep{Arnaud96} version 12.12.0, we simulate 1000 \xmm\ EPIC-pn spectra, with 100~ks exposure and a 0.5--10~keV flux of $10^{-11}$~erg~s$^{-1}$~cm$^{-2}$. We use the \textsc{relxill} model \citep[][]{Garcia14} to generate the spectra. We fix the outer emissivity index to 3, breaking to 2 inside 6 gravitational radii. This approximates the emissivity profile of a moderately extended corona \citep[e.g.][]{Wilkins12}, and means that the profile does not change drastically for different values of the spin. While steep emissivity profiles are commonly measured in AGN, these are only realistic in the case of high spin and we need a profile that is viable regardless of the spin parameter. We fix the photon index to 2, and the ionization to $\log(\xi)=1$. We then draw 1000 random parameter combinations for the other parameters, from uniform distributions within specified limits (Table~\ref{tab:ref_control_pars}). 
For each parameter combination we simulate a spectrum using \textsc{xspec} version 12.12.0 \citep[][]{Arnaud96}, using the \texttt{python} wrapper \textsc{pyxspec} to automate the process, and then bin that spectrum according to the scheme outlined by \citet{Kaastra16} using the \textsc{ftgrouppha} tool included as part of \textsc{heasoft} (version 6.29).

\begin{table}
    \centering
    \caption{Parameters used to simulate the reflection control spectra. We simulate spectra based on 1000 parameter combinations where each parameter is drawn randomly from a uniform distribution between the upper and lower bounds.}
    \begin{tabular}{l c c r}
    \hline \hline
         Parameter & Lower bound & Upper bound & Description\\
         \hline
         $q_\mathrm{in}$& 2$^a$ & & Inner emissivity index\\
         $q_\mathrm{out}$& 3$^a$ & & Outer emissivity index \\
         $r_\mathrm{break}$ & 6$^a$ & & Break radius ($r_\mathrm{G}$)\\
         $a$ & 0 & 0.98 & Spin\\
         $\cos(i)^b$ & 0.2 & 0.95 & Inclination\\
         $r_\mathrm{in}$ & $1^a$ & & Inner radius ($r_\mathrm{ISCO}$)\\
         $r_\mathrm{out}$ & $400^a$ & & Outer radius ($r_\mathrm{G}$)\\
         $\Gamma$ & 2$^a$ & & Photon index\\
         $\log(xi)$ & 1$^a$ & & Ionisation (erg~cm~s$^{-1}$)\\
         $A_\mathrm{Fe}$ & 1 & 5 & Iron abundance (solar)\\
         $E_\mathrm{cut}$ & 300$^*$ & & Cutoff energy (keV)\\
         $\cal{R}$ & 1 & 3 & Reflection fraction\\
         \hline
         \hline
    \end{tabular}
    $^a$These parameters are fixed in the simulations.\\
    $^b$We use cos($i$) for consistency with the disk wind model.\\
    \label{tab:ref_control_pars}
\end{table}

We then use \textsc{pyxspec} to fit each of the 1000 simulated reflection spectra from 3--10~keV with the same model, with the fit parameters initialised at the simulated values. We free the spin, inclination, iron abundance, reflection fraction, and photon index parameters, and then run the fit algorithm. In each case, we record the best fit value. For computational reasons, we do not calculate errors (it would drastically increase the run time, and the error algorithm frequently requires manual intervention). However, we test the errors in a few specific cases and find that they accurately reflect the scatter in the data.

\begin{figure*}
    \centering
    \includegraphics[width=0.49\linewidth]{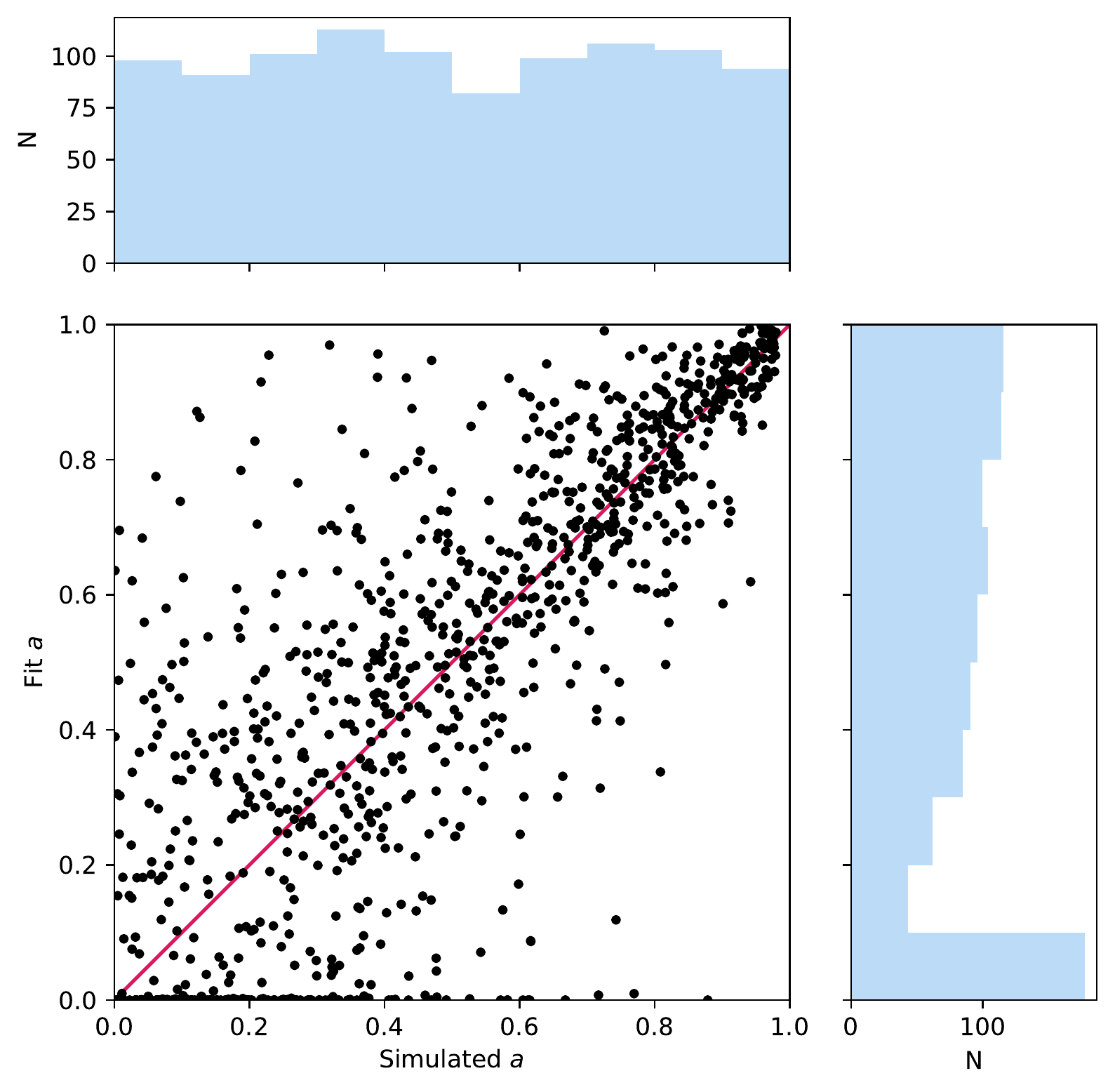}
    \includegraphics[width=0.49\linewidth]{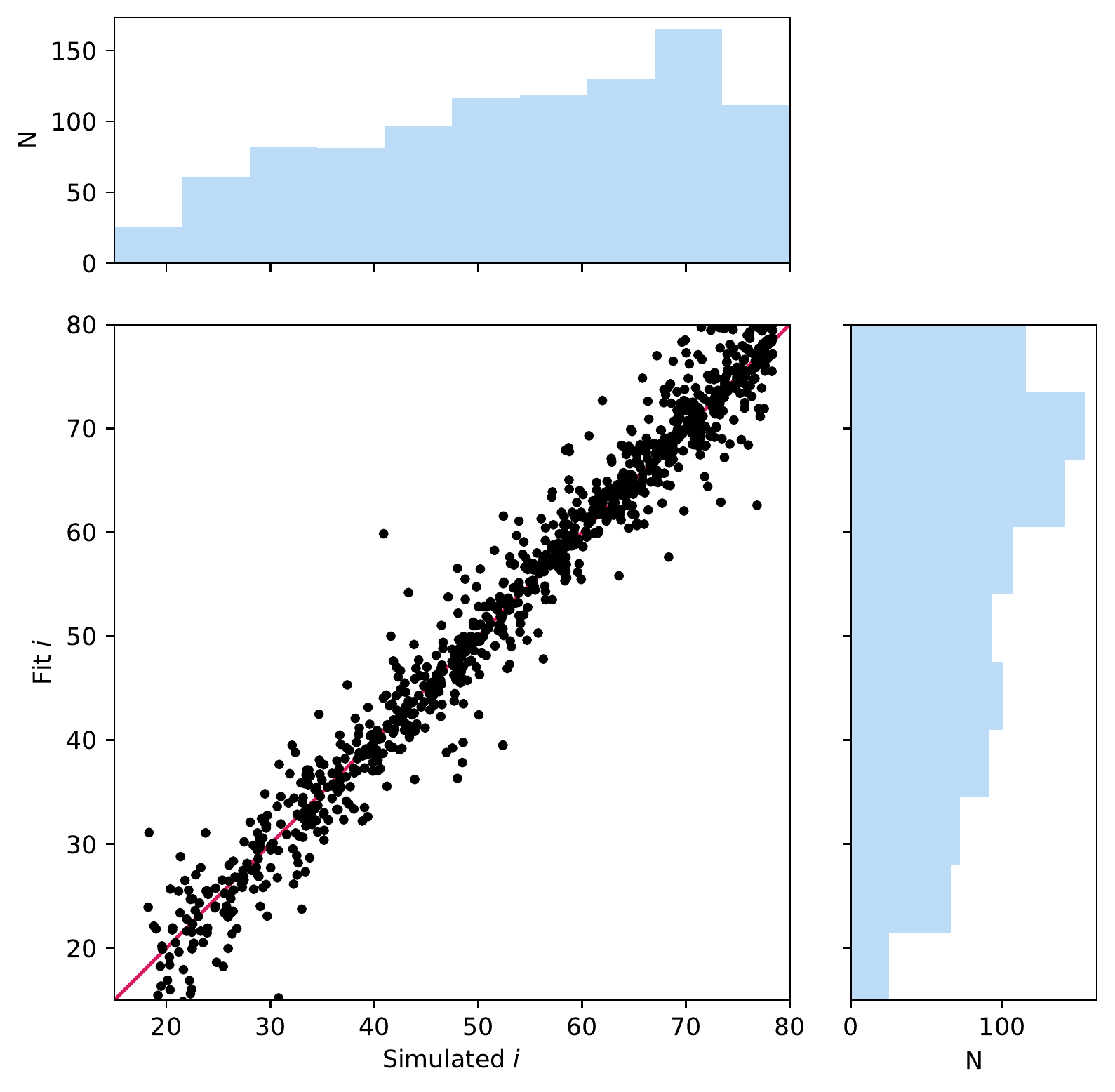}
    \caption{Parameter recovery for the reflection control simulations, where a pure reflection model is used to simulate \xmm\ spectra and the same model is used to fit the spectra. In each case the x axis shows the simulated value of the parameter (spin and inclination, respectively) and the y axis shows the value returned from spectral fitting. The histograms show the simulated and fit parameter distributions. At high spins, accurate values of spin are reliably returned. At lower spins, the accuracy decreases, but is not systematically offset. An excess of points appear at $a=0$, as we do not allow values of $a<0$. The inclination is reliably returned, regardless of the simulated value of $i$. The Pearson correlation coefficients are 0.79 and 0.98, respectively.}
    \label{fig:ref_control_recovery}
\end{figure*}

We then examine how well the parameters are recovered, relative to their simulated values. We focus on the spin and inclination, as they are arguably the most astrophysically important parameters. The recovery of these parameters is shown in Fig.~\ref{fig:ref_control_recovery}. The spin parameter is recovered well at high spin values, but significantly more scatter is introduced at lower spins. This is a well known feature of reflection spin measurements, as spin is much easier to measure when it is high \citep[see e.g.][]{Reynolds21}. The distribution of fit spin values matches the simulated distribution well, except for an excess at $a=0$ where the parameter space is artificially truncated. The inclination is well recovered, regardless of the simulated value, and the simulated and fit distributions match well. 

\subsubsection{Disk wind}

We use the same procedure as the reflection case to perform a set of control simulations. We draw 1000 parameter combinations uniformly from a pure disk wind model (\textsc{diskwind $\times$ powerlaw}; parameter ranges are given in Table~\ref{tab:dw_control_pars}), simulate the corresponding spectra, and then fit them with the same model, with parameters initialised at the simulated value.

\begin{table}
    \centering
    \caption{Parameters used to simulate the disk wind control spectra. We simulate spectra based on 1000 parameter combinations where each parameter is drawn randomly from a uniform distribution between the upper and lower bounds.}
    \begin{tabular}{l c c r}
    \hline \hline
         Parameter & Lower bound & Upper bound & Description\\
         \hline
         $\mu$ & 0.2 & 0.95 & Inclination\\
         $\dot{M}$ & 0.1 & 0.5 & Outflow rate ($\dot{M}_\mathrm{Edd}$)\\
         $f_\mathrm{v}$ & 0.25 & 1.5 & Velocity parameter\\
         $L_\mathrm{X}$ & 0.5 & 2 & Luminosity ($L_\mathrm{Edd}$)\\
         $\Gamma$ & 2$^a$ & & Photon index\\
         \hline
         \hline
    \end{tabular}
    $^a$These parameters are fixed in the simulations.\\
    \label{tab:dw_control_pars}
\end{table}

\begin{figure*}
    \centering
    \includegraphics[width=0.49\linewidth]{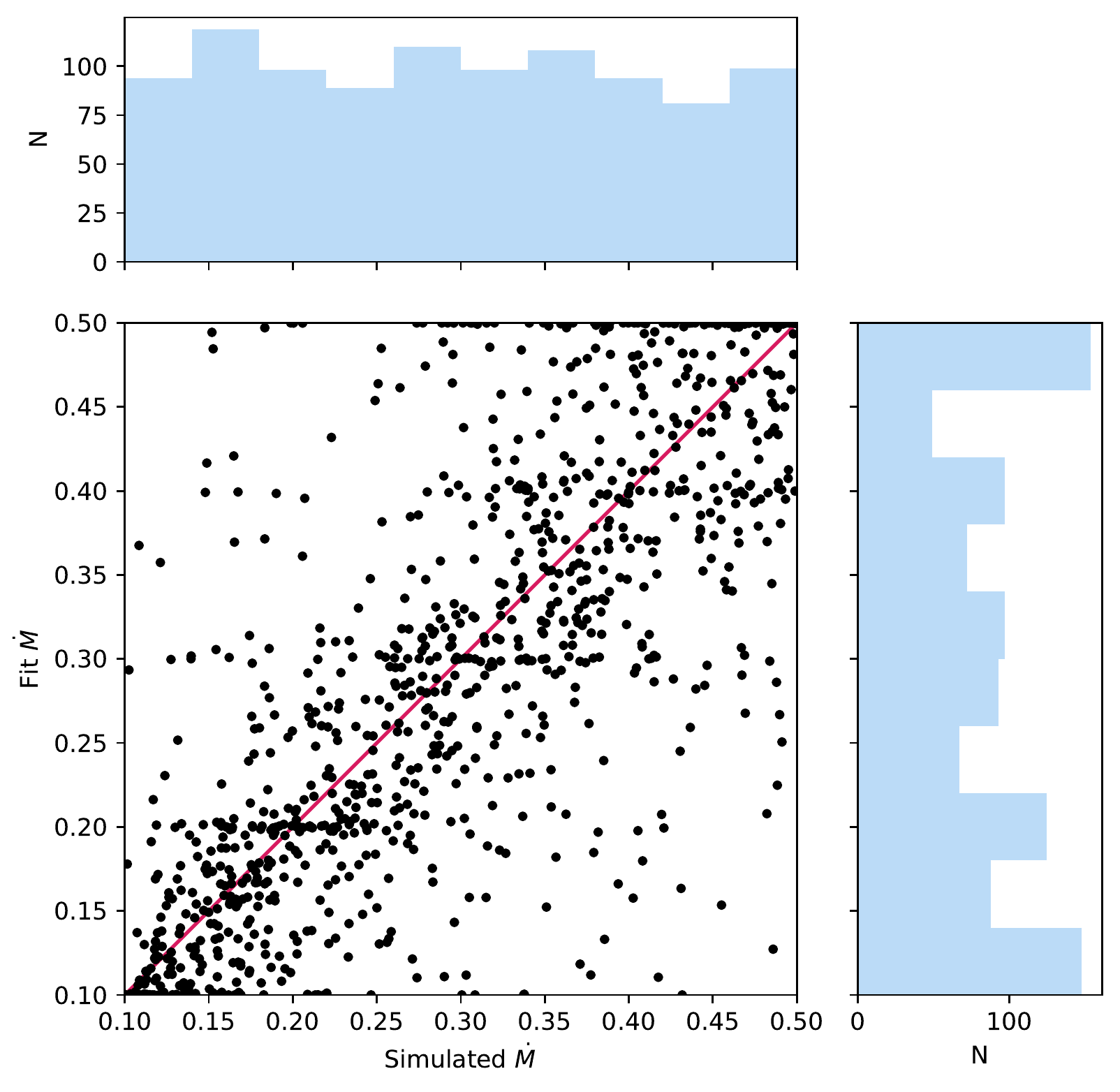}
    \includegraphics[width=0.49\linewidth]{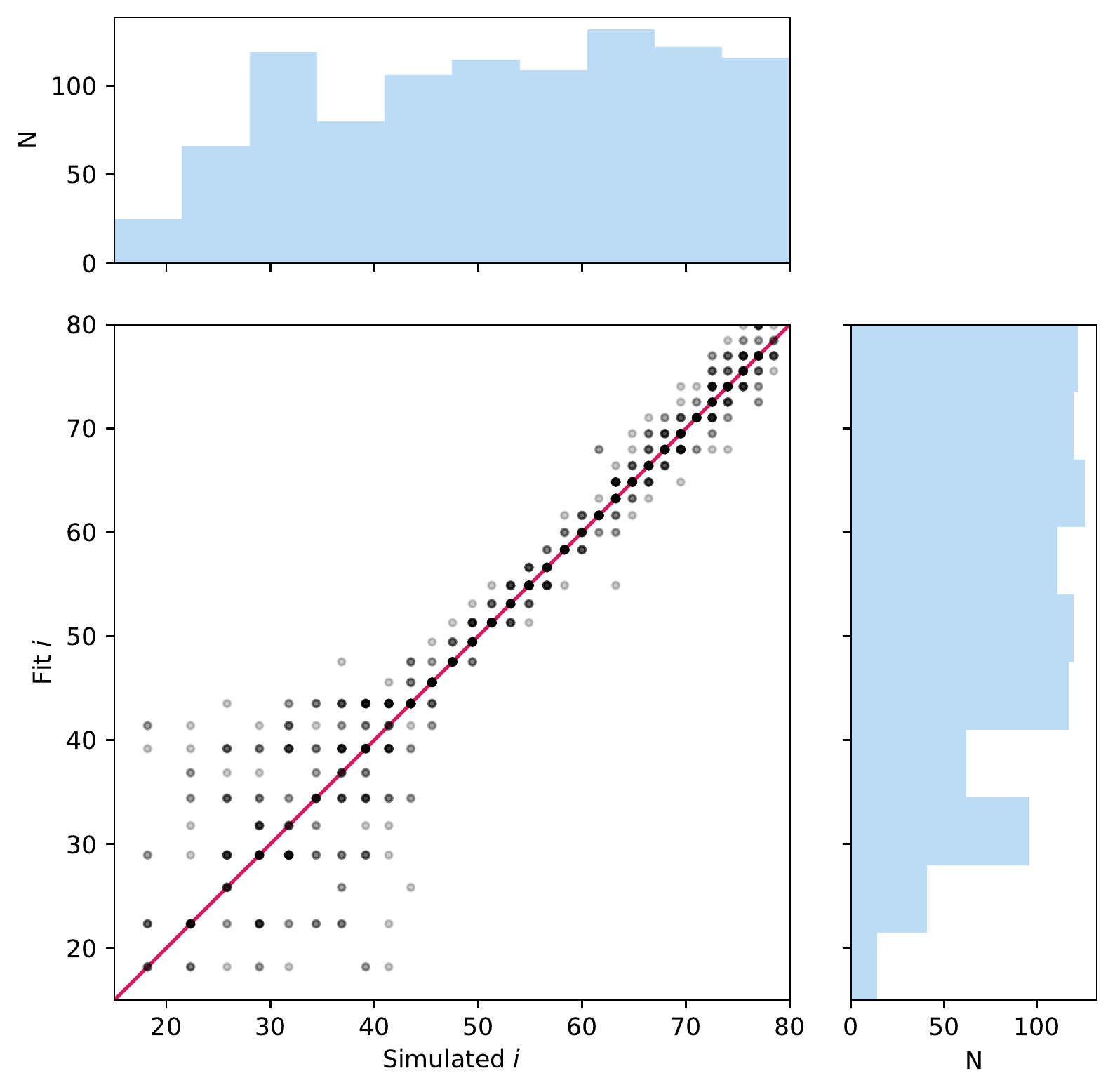}
    \caption{Parameter recovery for the disk wind control simulations, where a pure disk wind model is used to simulate \xmm\ spectra and the same model is fit to the simulations. In each case, the x axis shows the simulated parameter value, and the y axis shows the fit value. The left panel shows the mass outflow rate, which is recovered reasonably well, with a Pearson correlation coefficient of 0.76, although it tends to cluster at model grid points (every 0.1) due to imperfect interpolation within the grid. The right panel shows the inclination recovery, which is generally very good with a Pearson coefficient of 0.96. The scatter in the points increases below 45 degrees, when the line of sight no longer intercepts the wind, meaning no absorption is produced and the constraining power of the model is reduced. Note that during the simulations and fits the $\mu$ parameter is adjusted to the nearest grid point to avoid problems with the interpolation (see text), causing the points to be quantised in $\cos(i)$. To illustrate the density, the opacity corresponds to the number of fits at each grid point.}
    \label{fig:dw_control_recovery}
\end{figure*}

The fitting procedure differs slightly from the reflection case, as the $\mu$ and $f_v$ parameters do not interpolate well between gridpoints (this will be fixed in the newest version of the model; Matzeu et al., in prep). To work around this, in each case we run an initial fit with all the parameters free, then once a best fit has been obtained we fix the $\mu$ and $f_v$ to the nearest grid point (i.e. every $\Delta\mu=0.025$ and $\Delta f_v=0.25$). We then re-run the fit for the other parameters.

The results are qualitatively similar to the reflection fits. The parameters are generally well recovered, with some scatter in the fits. For simplicity, we focus on the mass outflow rate $\dot{M}$ and inclination $\mu$ parameters. The parameter recovery is shown in Fig.~\ref{fig:dw_control_recovery}. $\dot{M}$ is recovered reasonably well across the whole parameter space, with a slight excess in points at the edges of the parameter space and at grid points where the model is evaluated. The inclination recovery is better, particularly at high viewing angles. At angles below 45 degrees (the wind opening angle), the line of sight does not intersect the wind, weakening the constraint on $i$ noticeably but not introducing any systematic bias.

\subsection{Hybrid Model}
\label{sec:hybridsims}

Having tested the parameter recovery for the disk wind and reflection models in isolation, we now investigate how they operate when applied to a more complex spectrum. Specifically, we consider spectra where both disk reflection and winds are present simultaneously, and therefore both processes contribute to the net Fe~K emission profile. For this, we combine the two models discussed above, using \textsc{relxill} and the disk wind model together. In \textsc{xspec} format, the model used to simulate the spectra is \textsc{diskwind $\times$ relxill}. The parameters are the same as discussed above for the control samples, with the exception of the outer radius for the reflection model. We truncate the reflection at an outer radius of 32$r_\mathrm{G}$, the minimum launching radius of the wind. Additionally, we tie the reflection and disk wind inclinations together, so that the model is self-consistent. As with the control sample, we draw 1000 parameter combinations, and simulate the corresponding 100~ks \xmm\ EPIC-pn spectra. To explore the future impact of this problem, we also simulate 100~ks \athena\ X-ray Integral Field Unit (XIFU) spectra with the same parameter distributions, fluxes, and exposure times.

We next fit each simulated spectrum with the reflection and disk wind models \textit{separately}, using the same model set up as the control fits. This tells us how good the parameter recovery will be when a single process is assumed to dominate, if that is not actually the case.

\subsection{Reflection Parameter Recovery}

To fit the hybrid spectra, we first use the same reflection model as in Section~\ref{sec:refcontrol} but with the addition of two Gaussian absorption lines to account for any absorption features in the spectrum. This represents a scenario where the inner disk is viewed through a relativistic wind, but the wind does not contribute significantly to the net Fe~K emission. With real data, this is a fairly common model set-up \citep[e.g.][]{Parker17_nature, Parker18_iras13349, Kosec18, Kosec20, Walton19}, using either Gaussian lines or a photoionised absorption model with no emission counterpart. 

The fitting procedure is the same as with the control samples, but we add in an extra step to account for the absorption. We run an intial fit, with the parameters initialised from their simulated values, then run a blind line-scan over the range from 6.7~keV to the energy corresponding to the terminal velocity of the wind to find the main iron absorption line. The second Gaussian absorption line energy is kept at a fixed ratio of $E_2=1.16E_1$ (the ratio of the hydrogren like Ni to Fe lines) to the first, to account for any Ni absorption in the spectrum. 

The majority of the hybrid \xmm\ spectra are well fit with the pure reflection model (714 of the 1000 spectra have $\chi^2_\nu<1.5$, only 149 have $\chi^2_\nu>2$). We show some randomly selected well fit spectra in Fig.~\ref{fig:both_models_spectra}. Examples of poorly fit spectra are shown in Appendix~\ref{sec:fitplots}. The poor fits are typically caused either by a noticeably double-peaked line profile, absorption structure that the simple Gaussian absorption we use cannot accurately model, or a simple failure of the fitting algorithm. This failure occurs either when the \textsc{xspec} algorithm gets stuck in a false minimum, or when the line scan identifies the wrong feature (e.g. the Fe K edge), also causing the fit to occupy a false minimum. We examine this further in Section~\ref{sec:discuss_caveats} and conservatively estimate that $\sim10\%$ of our fits are classified as poor fits when they should instead be good fits. The cases where the reflection model is unable to fully describe the double peaked emission profile correspond to spectra where the receding side of the wind is close to perpendicular to the line of sight, while the approaching side is strongly blueshifted. These spectra could likely be modelled in practice with the addition of a second reflection component, attributed to more distant material in the outer disk. This would then provide a strong relatively narrow line at 6.4 keV, while the inner disk reflection would describe the blueshifted emission.
In general, there is nothing in the poor fit spectra that suggests they could not be well fit with a pure reflection model with minor tweaks, such as the addition of a distant reflection component or a more complex absorption model.
The well fit spectra cover a wide range of different line profiles, including those with strong absorption lines and spectra where the absorption is absent altogether. In general, it is not possible without prior knowledge of the simulations to determine the true contribution of reflection to the line profile from these fits.

\begin{figure*}
    \centering
    \includegraphics[width=0.8\linewidth]{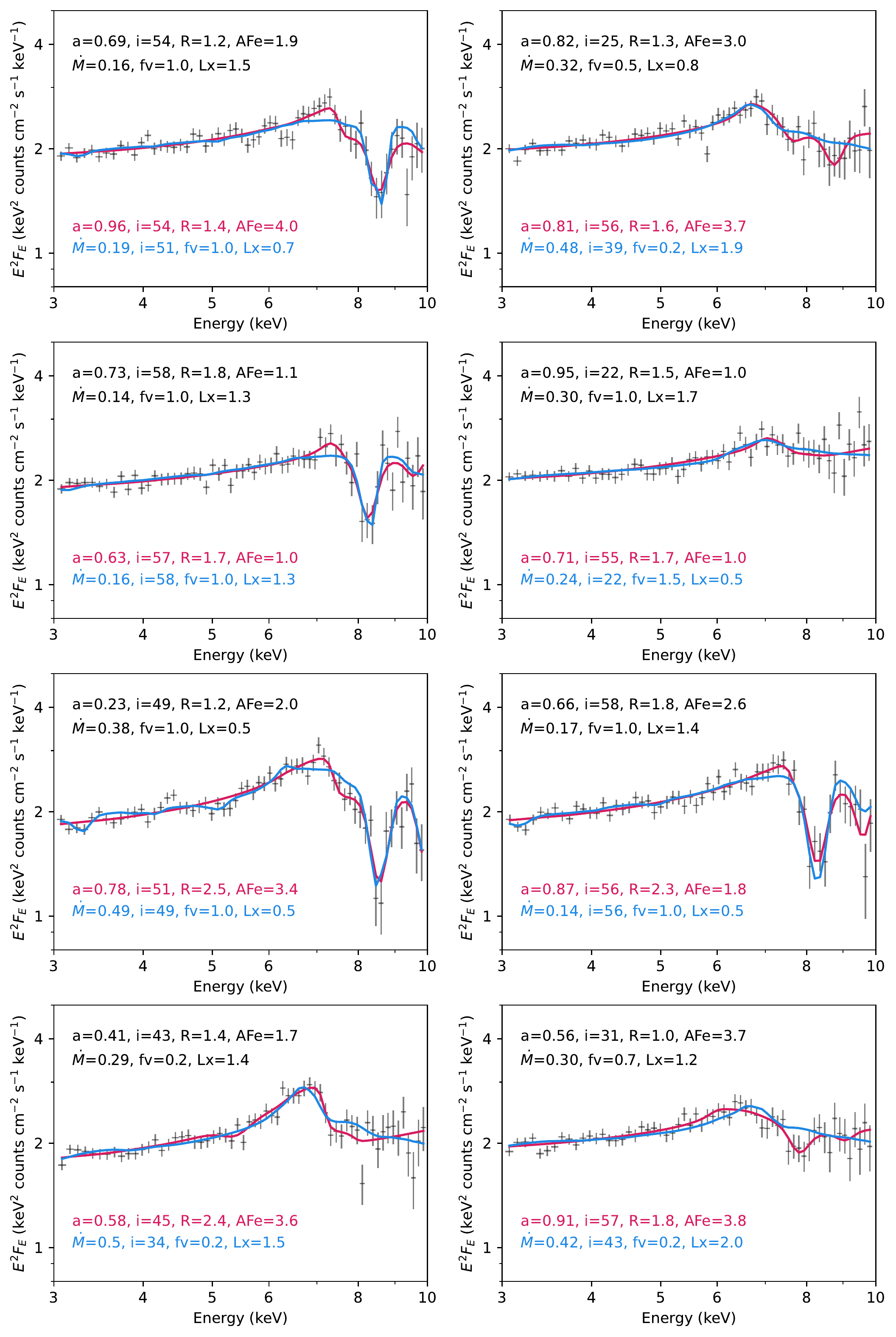}
    \caption{8 randomly selected \xmm\ spectra that are well fit with both the reflection (red) and disk wind (blue) models, with both models shown. Spectra are corrected for effective area, but not unfolded from the instrument response. In each case, the true simulated parameters are shown at the top, and the best fit parameters for each model are shown at the bottom. The models are extremely similar to each other, and cover a wide range of emission line profile shapes. Where differences exist between the models they are generally at high energies, where the spectrum is mainly determined by the wind absorption, rather than the emission line.}
    \label{fig:both_models_spectra}
\end{figure*}

\begin{figure*}
    \centering
    \includegraphics[width=0.49\linewidth]{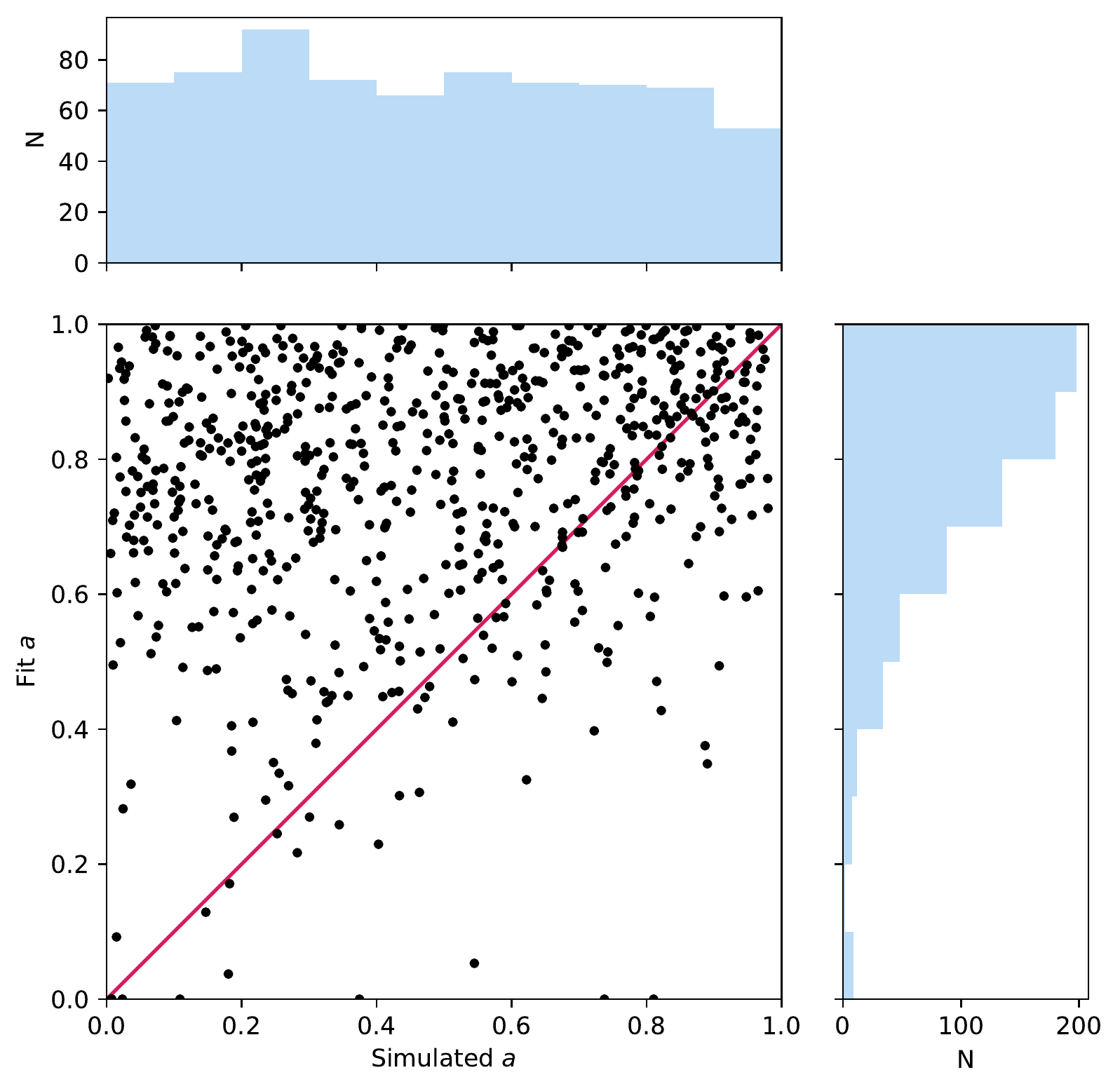}
    \includegraphics[width=0.49\linewidth]{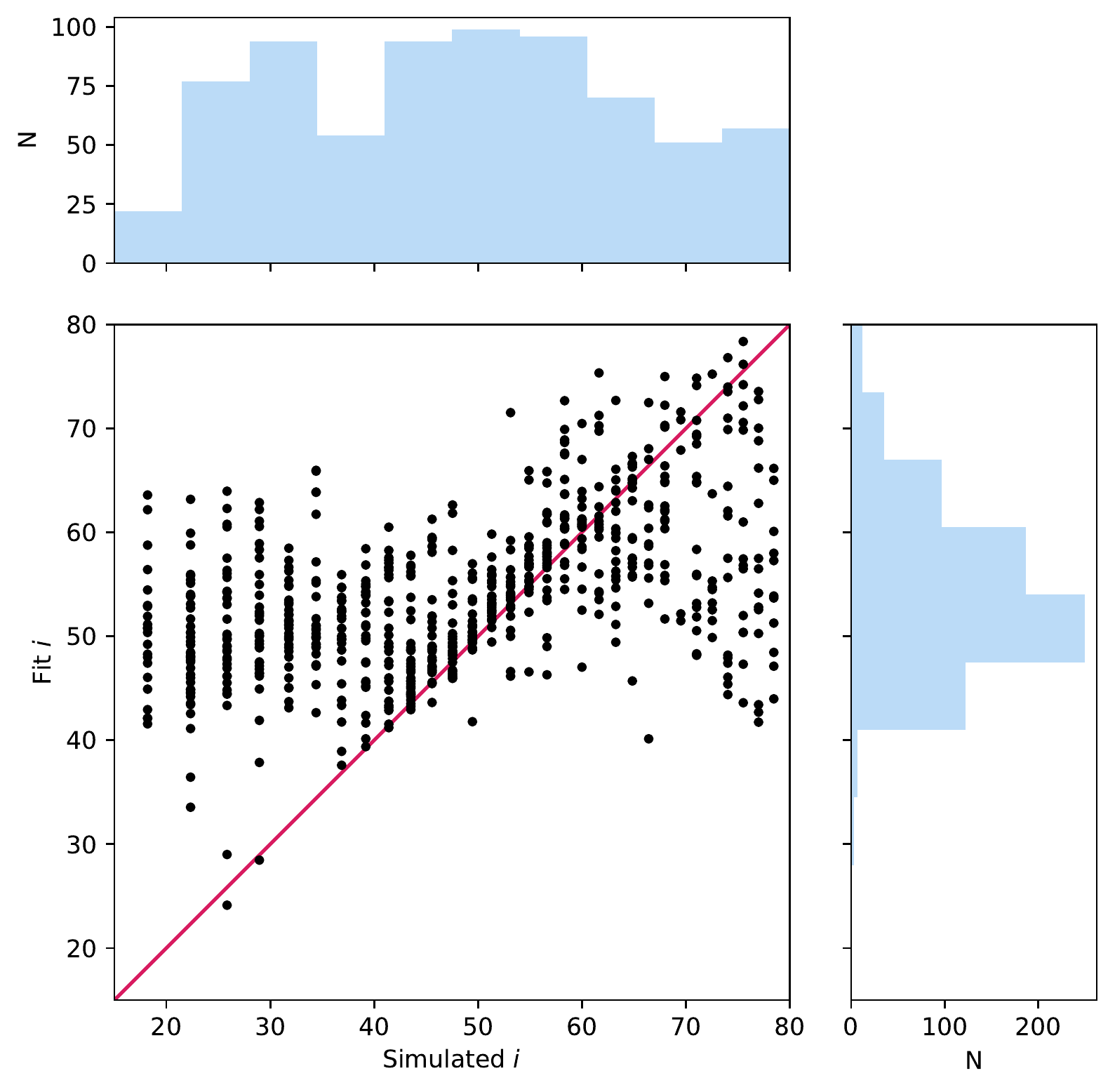}
    \caption{Parameter recovery for the pure reflection model, fit to hybrid reflection plus disk wind \xmm\ spectra. The spin recovery (left) is very poor, with a Pearson coefficient of 0.20. The estimated spin is strongly biased towards high spin values. The inclination recovery is also very bad, with a Pearson coefficient of 0.50. There is a strong bias towards intermediate values of inclination, around 50 degrees, and only simulated values close to this are recovered correctly.}
    \label{fig:ref_recovery}
\end{figure*}

The spin and inclination parameters recovered from these fits are shown in Fig.~\ref{fig:ref_recovery}. In contrast to the control simulations, where the parameters are well recovered with minimal bias, these posterior distributions show very strong systematic biases, and a very poor recovery of the input parameters. The fit returns high spin, regardless of the simulated value, and the inclination is strongly concentrated around 50 degrees.

Switching to higher quality \athena\ XIFU data does not solve this problem. In fact, a larger fraction of the spectra are formally well fit (930 of 1000 spectra have $\chi^2_\nu<1.5$, 31 have $\chi^2_\nu>2$). This is an issue with the simple fit statistic rather than the instrument, but worth noting regardless. The increase in effective area with \athena\ is much larger at low energies ($<7$~keV), and leads to a very large number of energy bins at these energies after binning. These low energy bins are consistently well fit, regardless of the model. This leads to very low overall $\chi^2$ values. 
We discuss this further in Appendix~\ref{sec:chi2}. For our purposes, a better metric for the accuracy of the fits can be obtained by looking at the fit statistic in specific energy bands. We calculate $\chi^2_\nu$ in the 3--5~keV (continuum), 5--7~keV (emission line) and 7--10~keV (absorption lines) bands. In the continuum band,  944 spectra have $\chi^2_\nu<1.5$, and the spectra that are not well fit are characterised by clear absorption lines from Si, Ar and Ca. 
The emission line band is similarly well fit, with 962 spectra with $\chi^2_\nu<1.5$. Since the broad emission lines are resolved already at CCD resolution, the dramatic improvement in energy resolution with \athena\ does not break the degeneracy between the two models. We show some randomly selected well fit example spectra in Fig.~\ref{fig:both_models_spectra_athena}, and some poor fits are shown in Appendix~\ref{sec:fitplots}. 
The absorption band is more complex, and as expected has fewer good fits. Only 669 spectra have $\chi^2_\nu<1.5$, primarily because the simple double Gaussian absorption model we use does not give a full description of the absorption spectrum in this band. These features are resolved with \athena\ so this occurs more frequently that with \xmm\ spectra. However, in practise this would be addressed by using a more sophisticated absorption model (this would be computationally expensive and is not necessary for this work, as sufficient spectra are well fit for us to draw robust conclusions).

\begin{figure*}
    \centering
    \includegraphics[width=0.8\linewidth]{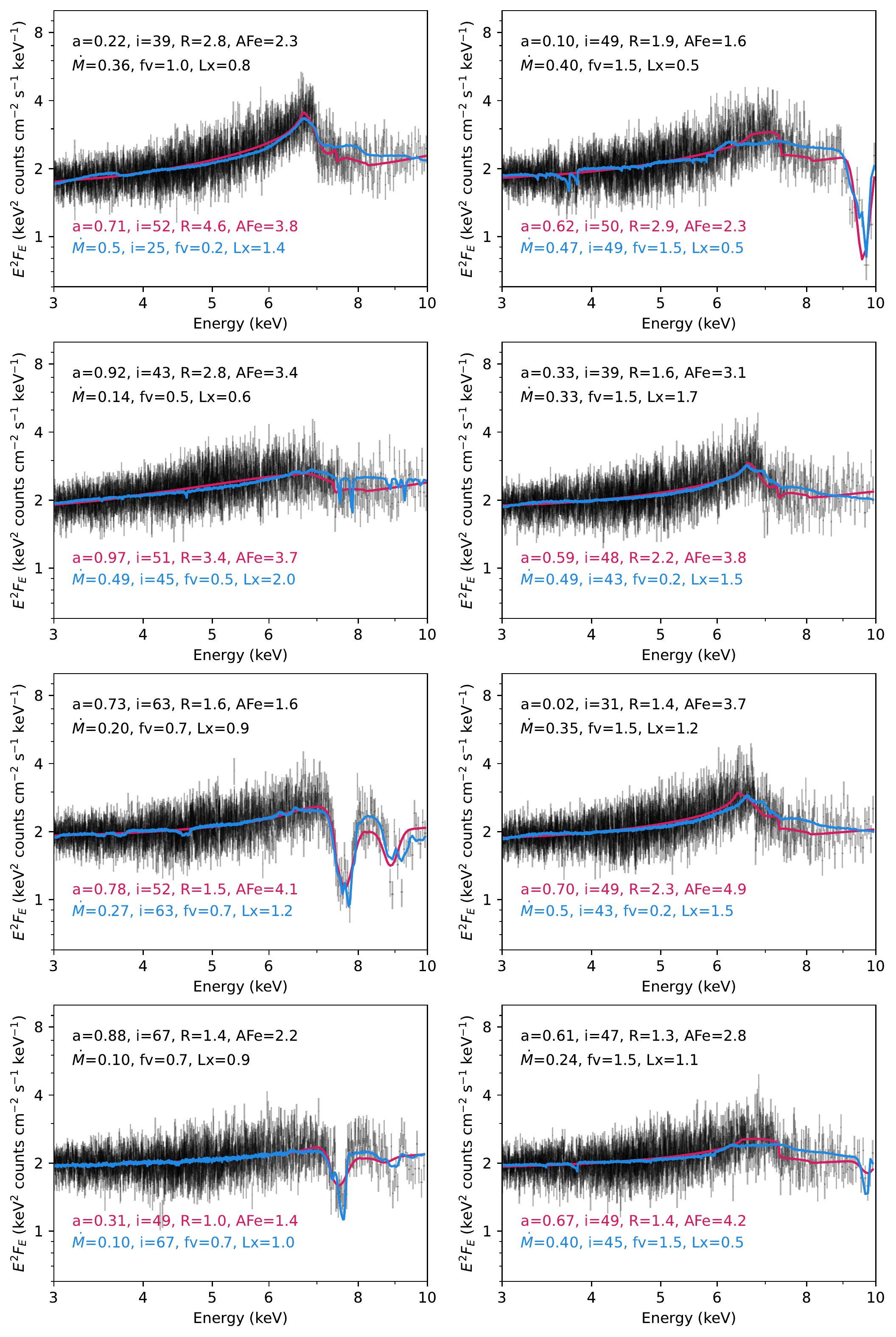}
    \caption{8 randomly selected \athena\ spectra that are well fit with both the reflection (red) and disk wind (blue) models, with both models shown. Spectra are corrected for effective area, but not unfolded from the instrument response. In each case, the true simulated parameters are shown at the top, and the best fit parameters for each model are shown at the bottom.}
    \label{fig:both_models_spectra_athena}
\end{figure*}

\begin{figure*}
    \centering
    \includegraphics[width=0.49\linewidth]{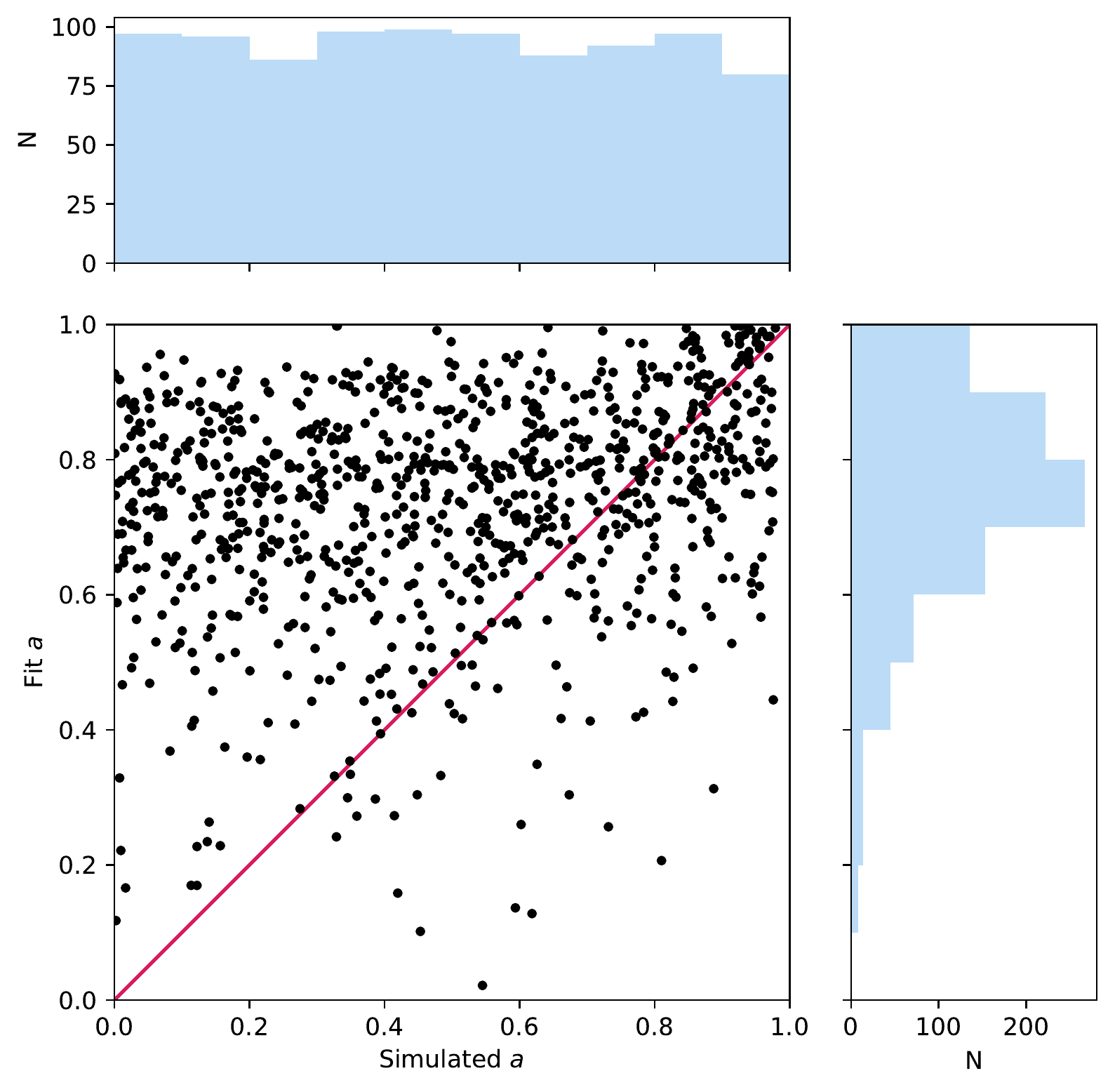}
    \includegraphics[width=0.49\linewidth]{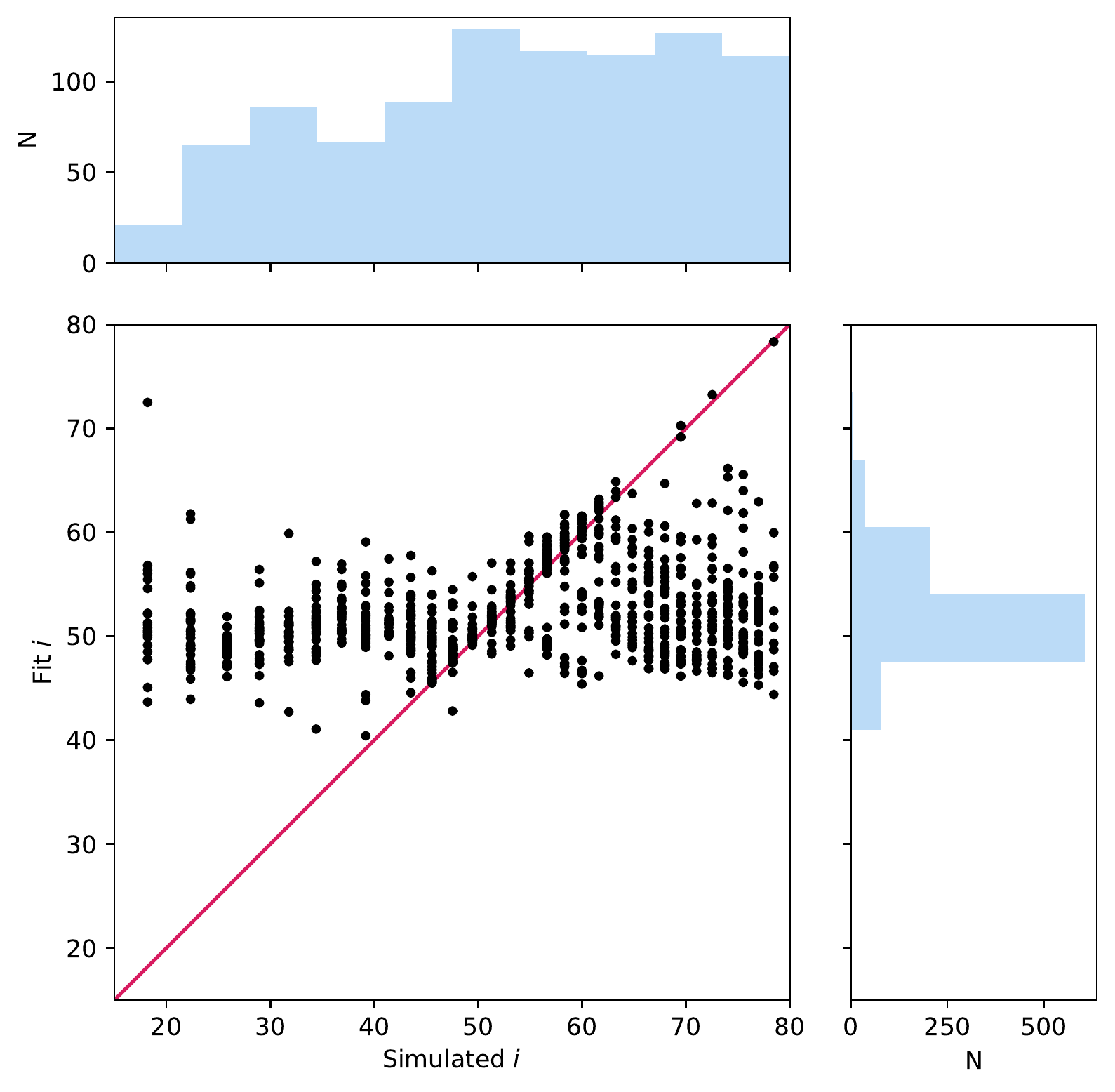}
    \caption{Parameter recovery for the reflection model, fit to hybrid reflection plus disk wind \athena\ XIFU spectra. This result is much the same as the \xmm\ case, but with less scatter. The spin is sill strongly biased towards high values, but with maximal spin excluded, and the inclination is strongly constrained to $\sim50$ degrees in all cases. The correlation coefficients are 0.21 and 0.23, respectively. }
    \label{fig:ref_recovery_athena}
\end{figure*}

The parameter recovery (shown in Fig.~\ref{fig:ref_recovery_athena}) is much the same as the \xmm\ case, but with a smaller scatter. The spin is still biased high, but is more concentrated on a value of 0.8. The inclination is very strongly concentrated on 50 degrees. We note that neither of these results should be taken as a general rule for how the disk/wind emission degeneracy will manifest, as they are likely specific to this particular model set up and assumed geometry.

Overall, we conclude that in the majority of cases a reflection model can easily fit a hybrid reflection plus wind spectrum, in which case the parameters returned from the fit will not be reflective of the true parameters. Importantly, without prior knowledge of the relative contributions from the two processes, it is likely impossible to determine whether this is happening purely from time-averaged X-ray spectroscopy.

\subsection{Disk Wind Parameter Recovery}

We next fit the hybrid spectra with the pure disk wind model, again using the same procedure as with the control sample. 
Similarly to the reflection case, the majority of the hybrid spectra are well fit with a pure disk wind model. 627 of the 1000 \xmm\ spectra have $\chi^2_\nu<1.5$, and 168 have $\chi^2_\nu>2$. 420 of the well fit spectra are also well fit by the pure reflection model. A selection of poorly fit spectra are shown in Appendix~\ref{sec:fitplots}, and in Fig.~\ref{fig:both_models_spectra} we show a selection of spectra that are well fit.

Again, the well fit spectra cover a wide range of line profiles, both with and without clear absorption features. As expected, the wind model does a better job of fitting the absorption than the simple phenomenological approach we used for reflection. The cases where the model does not fit are typically those where the line profile is red- or blue-shifted but still sharply peaked. The disk wind model typically only produces a sharp peak at 6.4~keV. We stress that this is unlikely to be a general property of wind models, and is likely specific to this particular geometry/model setup.

\begin{figure*}
    \centering
    \includegraphics[width=0.49\linewidth]{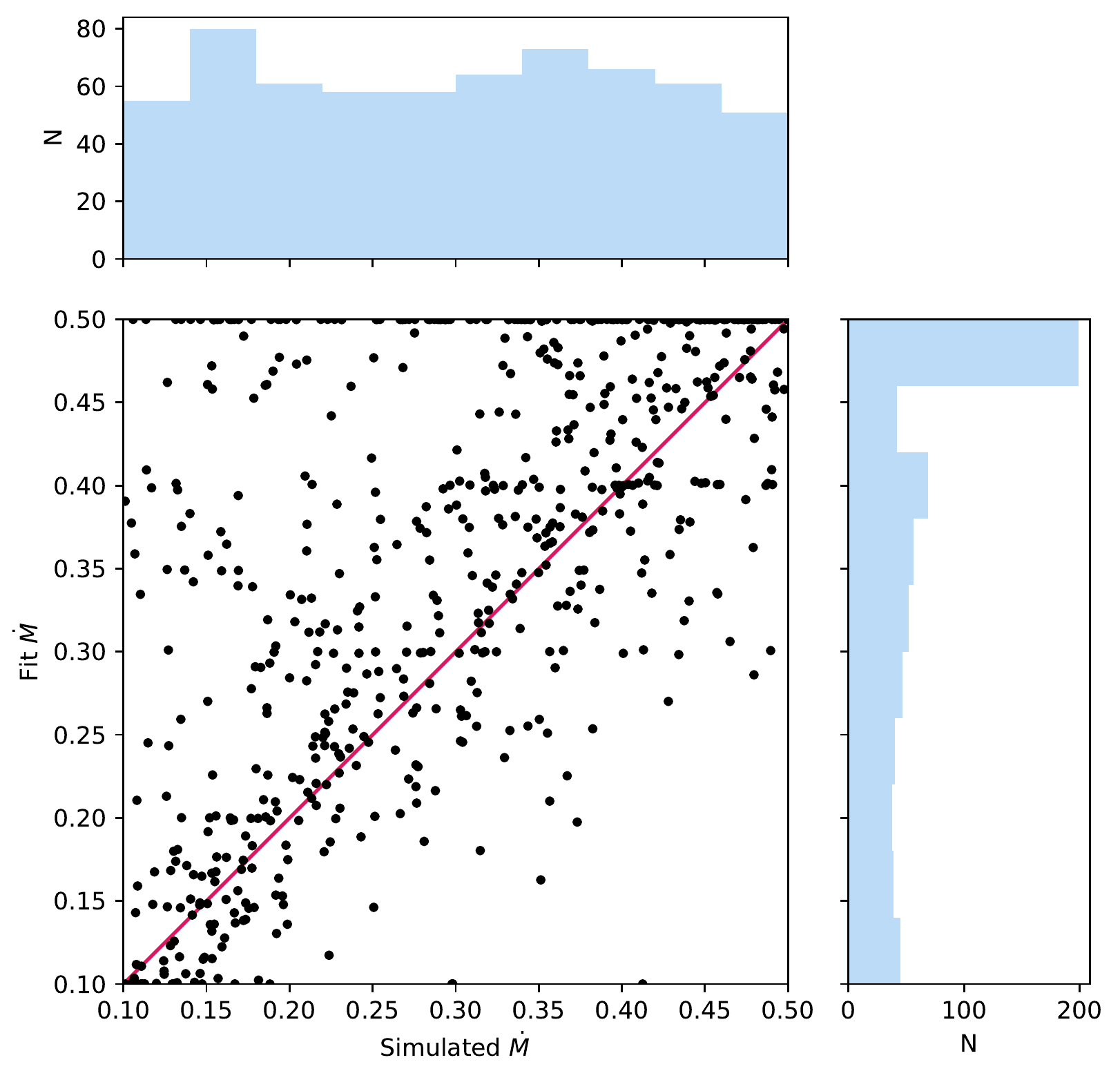}
    \includegraphics[width=0.49\linewidth]{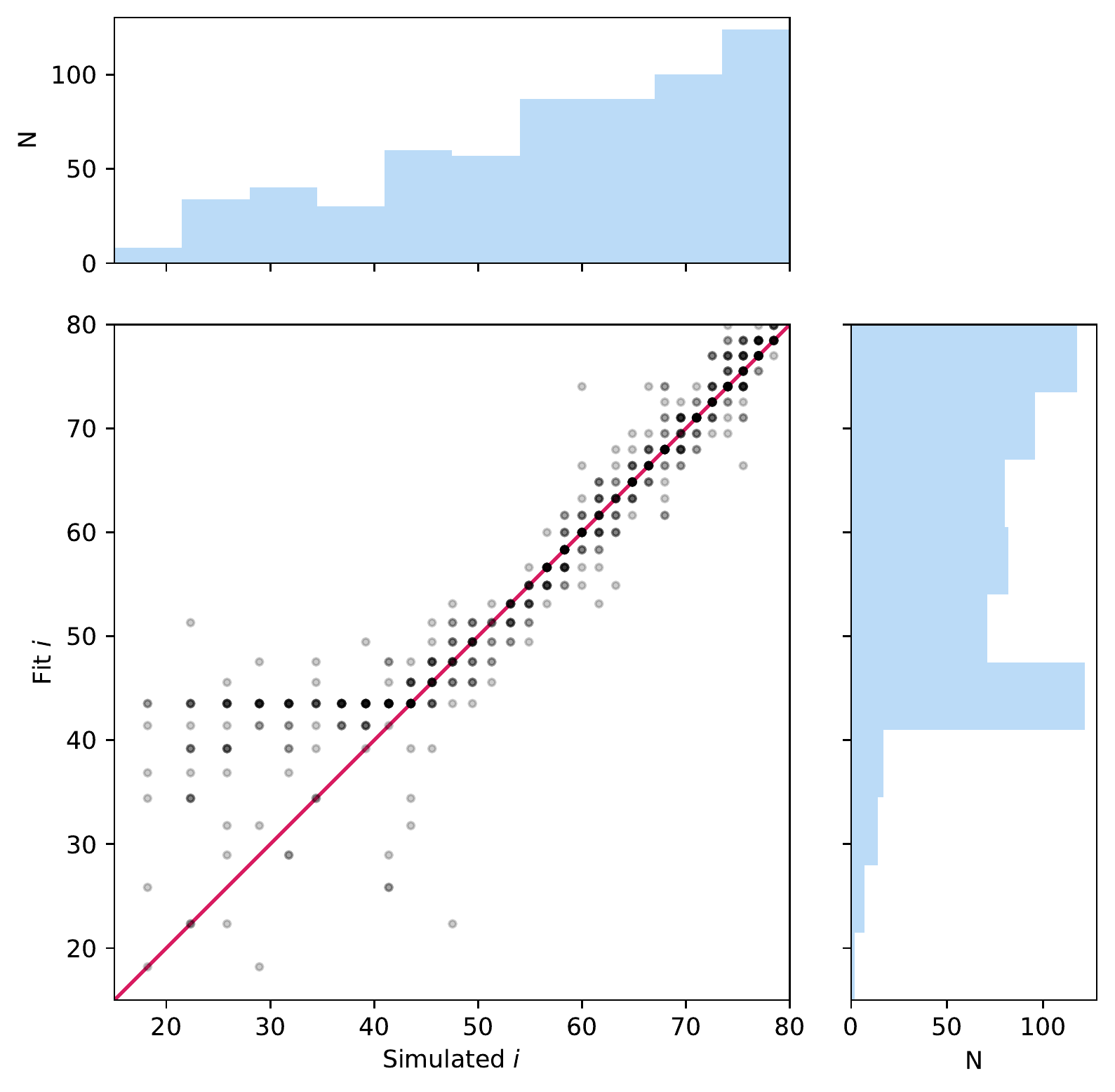}
    \caption{Parameter recovery for the disk wind model, fit to hybrid reflection plus disk wind \xmm\ spectra. The mass outflow rate recovery (left) is much worse than the control case (with a Pearson coefficient of 0.59), as a large fraction of the fits are at the upper limit, regardless of the simulated value. The case for inclination is better (Pearson coefficient of 0.93), as the inclination is reliably returned so long as the wind intersects the line of sight ($i>45$~degrees). Below 45~degrees, the measured inclination is independent of the simulated value. As with Fig.~\ref{fig:dw_control_recovery}, both simulated and fit values of $\mu=\cos(i)$ are adjusted to the nearest gridpoint, and we show the number of fits at each grid point with the opacity.}
    \label{fig:dw_recovery}
\end{figure*}

The parameter recovery for the inclination and mass outflow rates are shown in Fig.~\ref{fig:dw_recovery}. The biases introduced are less extreme than in the reflection case, but still significant. The mass outflow rate recovery superficially reasonable, but a very large fraction of the returned values are found at the upper limit of the parameter space. This higher mass outflow rate is likely necessary to account for the excess emission from reflection. The inclination is well recovered, so long as the line of sight intersects the wind. Below 45 degrees the recovered inclination is largely independent of the simulated value, giving the same value (just below the wind opening angle) in most cases. This is because no absorption feature is present for these angles, removing one of the key constraints. This is the case for most parameters, which are much less reliably constrained when no absorption line is present and the constraints must be derived from the emission line instead.

\begin{figure*}
    \centering
    \includegraphics[width=0.49\linewidth]{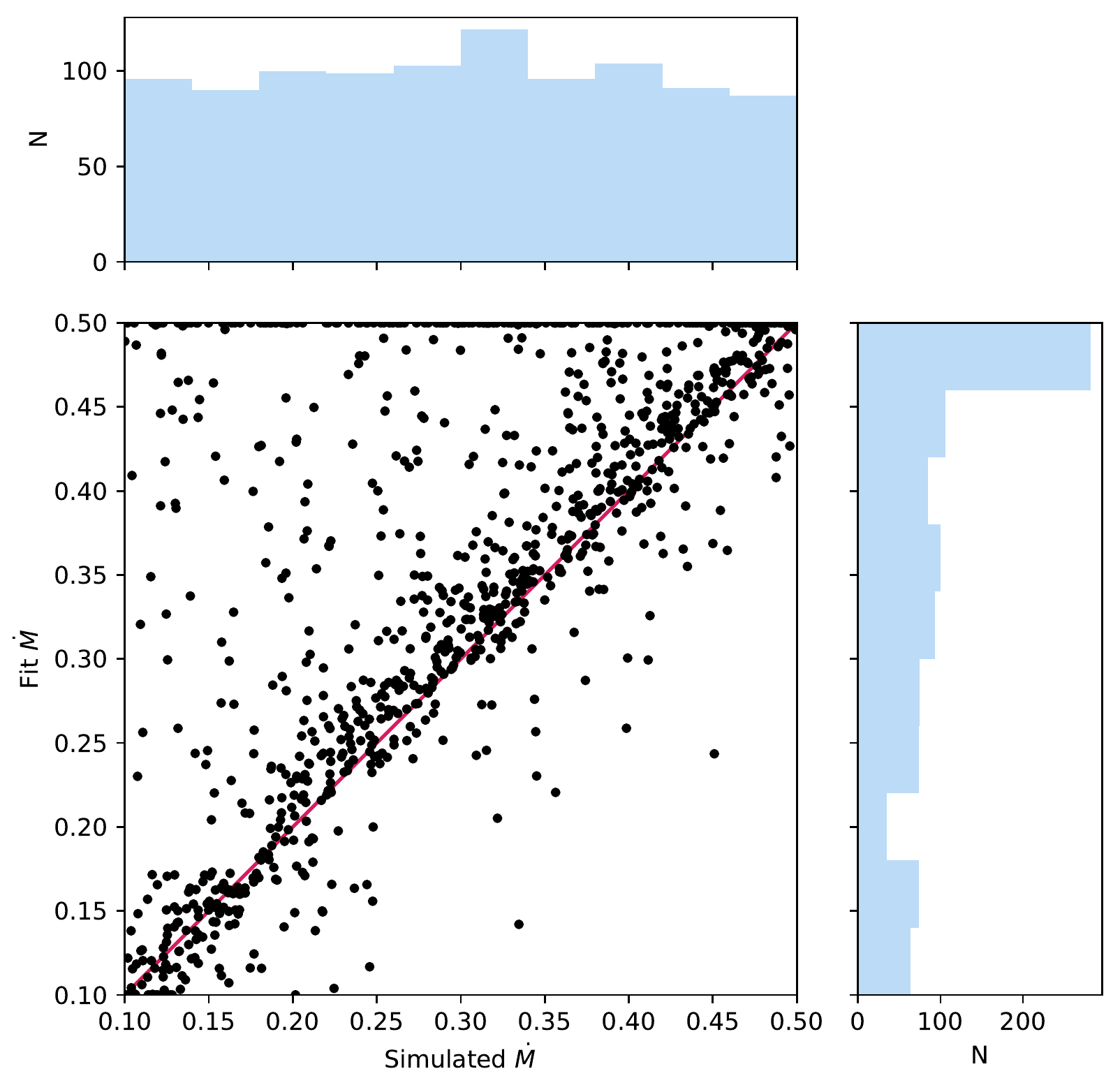}
    \includegraphics[width=0.49\linewidth]{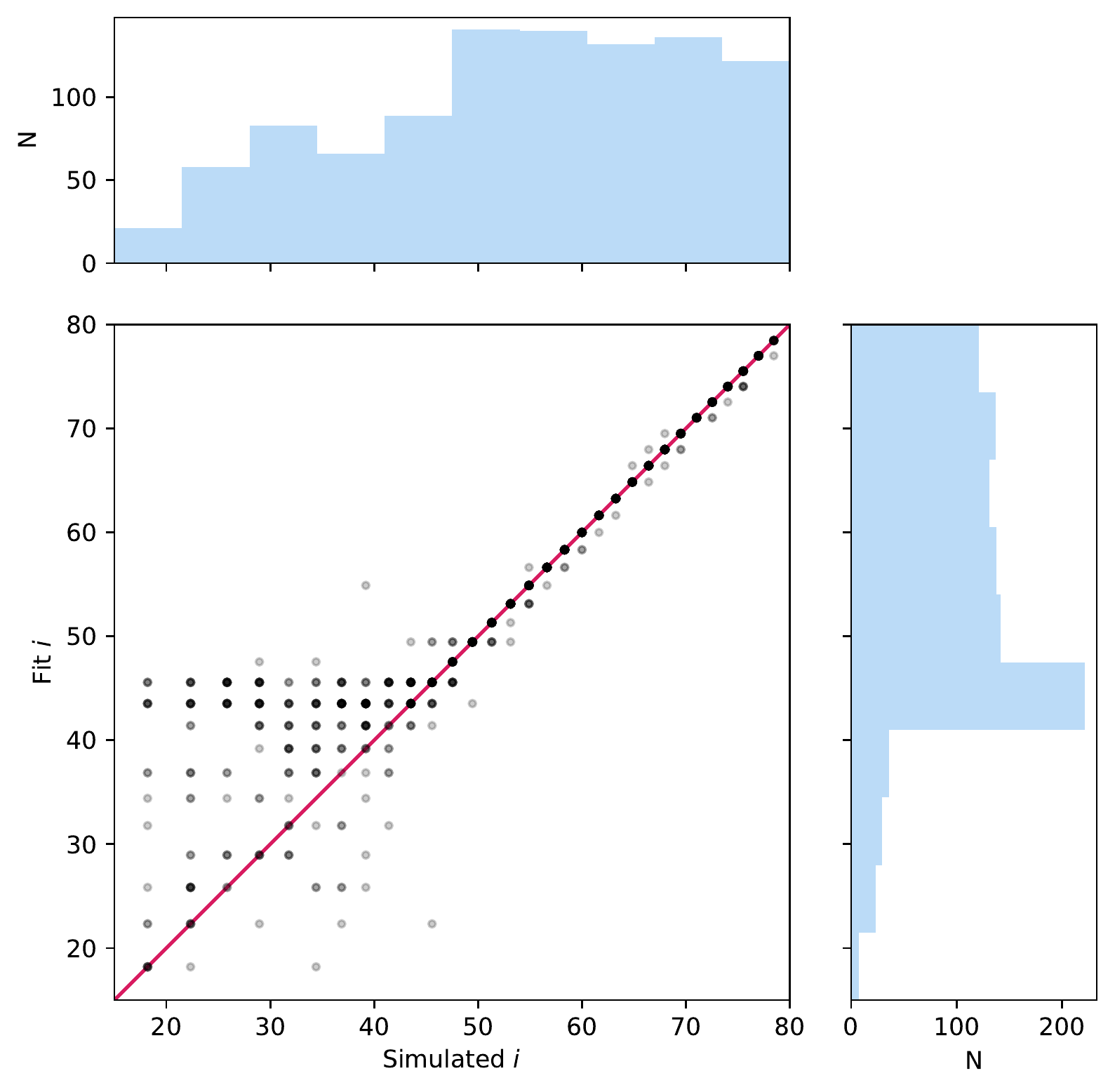}
    \caption{Parameter recovery for the disk wind model, fit to hybrid reflection plus disk wind \athena\ XIFU spectra. The results are qualitatively similar to the \xmm\ case, with the mass outflow rate still strongly biased towards higher values, and the inclination only reliable when the wind crosses the line of sight. The correlation coefficients are 0.66 and 0.94, respectively. As with Fig.~\ref{fig:dw_control_recovery}, both simulated and fit values of $\mu=\cos(i)$ are adjusted to the nearest gridpoint, and we show the number of fits at each grid point with the opacity.}
    \label{fig:dw_recovery_athena}
\end{figure*}

The same effect seen with the reflection spectra is evident when moving to the \athena\ XIFU data - the higher number of bins, preferentially at lower energies, leads to an improved fit statistic. As a result, 993 of the spectra have $\chi^2_\nu<1.5$, and 0 have $\chi^2_\nu>2$. 
Again, a clearer picture can be obtained from the energy resolved $\chi^2_\nu$. None of the spectra have a poor fit ($\chi^2_\nu>1.5$) in the continuum band (3--5~keV), as the disk wind model reliably fits any absorption features that appear here. Similarly to the reflection case, the vast majority of the iron line profiles are well modelled, with 974 spectra having $\chi^2_\nu<1.5$. Finally, the absorption band is the least well fit, with 708 spectra with $\chi^2_\nu<1.5$. This is also very similar to the reflection case, despite the more sophisticated absorption spectrum included in the disk wind model. The likely cause of this is the fit prioritising the higher signal lower energy bands, which dominate the overal $\chi^2$, at the expense of the lower signal high energy band. In practice, this could likely be accommodated in the modelling by decoupling the absorption and emission spectra, or adding additional wind layers, without having to include a reflection component.
The same systematics are visible in the parameter recovery with the \athena\ data as with the \xmm\ data (see Fig.~\ref{fig:dw_recovery_athena}). A large fraction of the mass outflow rates hit the upper limit, and the inclination is unreliable below the opening angle of the wind.

The results of this test are qualitatively similar to the reflection case, in that including an unmodelled reflection component introduces a large systematic error into the recovery of key parameters. We note that the bias introduced is smaller here than in the reflection case, but we stress that this is likely due to the model set up rather than an intrinsic property of the data. In particular, we note that the disk wind model has fewer free parameters than the reflection model, effectively meaning it is more constrained prior to fitting, and that much of the constraining power is due to the simultaneous fitting of the emission and absorption features. We note that short timescale variability in the absorption feature due to microstructure in the wind or response to the continuum may mean that this is not always a reliable approach.

\section{Discussion}
\label{sec:discussion}

\subsection{Implications for disk wind modelling}
\label{sec:discuss_diskwind}

A common aim of UFO spectroscopy is to understand the impact of UFOs on AGN feedback. To measure the feedback potential of a UFO, we need to know the geometry and kinetic power of the wind. Some constraints can be derived from UFO absorption lines, however these typically leave at least an order of magnitude of uncertainty in the relevant quantities. UFO absorption spectroscopy is fundamentally limited for two reasons: firstly, it only samples material along the line of sight, so it returns very little information about the overall geometry, and secondly because it only measures the integrated material along the line of sight (i.e. the column density, rather than the true density), leaving a huge degeneracy between the density and extent of the wind along the line of sight. Emission spectroscopy of the wind offers a path to circumvent these limitations, because the emission profile samples the whole X-ray illuminated wind volume, and encodes information about the geometry.

The degeneracy we have identified here means that it is impossible to derive reliable constraints from the emission part of the wind without having strong prior constraints on the reflected emission. Because in a disk wind scenario a disk is by definition present, the emission from the disk must always be accounted for, regardless of whether it offers a statistical improvement to the fit. Results derived from modelling of P-Cygni profiles in AGN that do not take into account the disk will be systematically biased, with the systematic error potentially much larger than the true signal. 

The main advantage for disk wind modelling, relative to modelling the disk emission, is the presence of the absorption line (note that the systematics dramatically worsen in our fits when the line of sight does not intersect the wind). The absorption line acts as a strong constraint on certain parameters, such as the ionisation state of the wind and the final velocity, restricting the parameter space available to the model. However, we note that this may be exaggerated in our model set up, as the behaviour of the absorption feature on the timescales of individual observations may not be representative of the global wind properties. The absorption may instead be showing local properties of clumps in the wind, so might not be as viable as a means of constraining the wind geometry as we have assumed. This may be more of an issue for higher mass AGN (such as PDS~456) where the relevant timescales are longer, while in lower mass AGN the variability in the absorption induced by small scale structure may be averaged over on observational timescales.

\subsection{Implications for reflection modelling}
\label{sec:discuss_reflection}

This degeneracy is clearly a major problem for reflection measurements in AGN. The advantage for reflection spectroscopy over disk wind spectroscopy is that reflection can in principle exist without the presence of a disk wind, while a disk wind by definition requires a disk. We also see clear evidence for reflection in X-ray binaries, where no UFOs have been observed, and microlensing results suggest that at least some of the relativistic Fe~K$\alpha$ emitting region must be compact, at least in some cases \citep[e.g.][]{Chartas12}. However, it is likely impossible to demonstrate from X-ray spectroscopy alone that a powerful disk wind is not present in any given observation of an AGN, out of the line of sight, and this can introduce a huge systematic error into key parameters measured with reflection spectroscopy.

This is particularly important for the black hole spin, usually the most sought after reflection parameter. The spin acts as a probe of the black hole formation and growth history, so reliable measurements of spin are crucial. Our modelling suggests that the presence of a disk wind introduces an overwhelming bias in spin measurements towards high spin.

It is also possible that the additional Fe~K emission from the disk wind contributes to the high Fe abundances frequently measured in AGN. This is unlikely to be the primary cause, as the same effect is also seen in XRBs without relativistic winds \citep[see e.g.][]{Tomsick18,Jiang19_gx339}, but any enhancement of the Fe emission relative to other reflection features should lead to a higher inferred abundance. This is difficult to test directly with out simulations, as the iron abundance is not generally well constrained in our fits as there are no other features to establish its relative strength. Many of the fits hit the upper and lower limits on the iron abundance in our control simulations due to the poor constraints offered by narrow band fitting. More fits hit the upper limit in the test simulations, suggesting that a bias towards higher abundances is introduced, but this could be due to the increased uncertainty in the fits and further work is needed to establish if this effect is genuine.

\subsection{Implications for Athena and XRISM}
\label{sec:discuss_xrism}

\citet{Barret19} use simulations to study ability of the \athena\ XIFU to measure black hole spin from reflection and simultaneously constrain absorption from warm absorbers and UFOs. In particular, it cleanly resolves absorption lines that overlap with the Fe~K line, a potential problem at CCD resolution \citep[][]{Middleton16}. However, they did not consider UFO emission in their simulations.

While it is possible that the improved energy resolution offered by the microcalorimeters on \xrism\ and \athena\ will somehow break this degeneracy, it is likely that the degree of velocity broadening involved in both emission mechanisms means that there are no narrow features to be resolved. In our simulations, the emission spectra from \athena\ offer no significant difference in reliability for distinguishing the emission mechanism or for parameter estimation. In fact, counter intuitively, the higher number of bins in the \athena\ spectrum means that the $\chi^2_\nu$ values are typically lower than for an equivalent \xmm\ fit, potentially hiding models that fit poorly at high energies. 

The main improvement of \athena\ over \xmm\ in our simulations typically lies in the absorption spectra, where in many cases lines are resolved by \athena\ that would not be visible at CCD resolution. Microcalorimeters will clearly be superb tools for studying the wind microstructure through absorption lines, but generally less revolutionary for studying broader features where energy resolution is not the limiting factor. We conclude that the wind/reflection degeneracy will most likely not be broken by \xrism\ or \athena , and will be a major problem for achieving some of their key science goals. To make use of these new instruments it is therefore necessary that we develop techniques to mitigate these effects in advance, rather than assuming that the problem will disappear with superior instrumentation.

\subsection{Mitigation strategies}
\label{sec:discuss_mitigation}

We have demonstrated that current modelling strategies of assuming that only relativistic reflection or disk wind emission contribute to the Fe~K emission line introduce huge systematic error, potentially making measurements of key parameters like the mass outflow rate or black hole spin impossible. Here, we discuss how this problem might be solved or mitigated.

Investigating these possibilities in detail is beyond the scope of this work, but we intend to follow up some of them in future work, as we encourage others to look into them as well.

\subsubsection{Broad-band spectroscopy}
\label{sec:broadband}
Our modelling has focused on the Fe~K band, but disk winds and reflection should leave subtler signatures outside this part of the spectrum. Using additional information from these regions will give some constraints on the relative contribution of the wind and disk emission.

The Compton hump, seen around 20--30~keV and frequently measured by \nustar , is a common signature of reflection from a dense accretion disk \citep[e.g.][]{Risaliti13}. Winds typically produce a weaker feature \citep[][]{Sim10, Nardini15}, so the presence of a strong Compton hump in excess of that predicted by distant reflection from the outer disk argues that a relativistic reflection component must be present, and gives some constraint on the Fe~K band contribution. We note that the relative strength of the Compton hump and Fe~K line depends on multiple parameters, most notably the iron abundance, so this constraint will not be perfect. The converse (that a weak or absent reflection hump means little or no contribution to the iron line from reflection) is not necessarily true. We have shown in Fig.~\ref{fig:pds} that the weak Compton hump in PDS~456 can easily be accommodated by a reflection model.

As a simple test of this, we consider the 2013 \nustar\ spectrum of Mrk~335, which has a strong Compton hump and is consistent with being in a reflection dominated state \citep[][]{Parker14_mrk335}. We fit the time-averaged spectrum with a hybrid model, including a disk wind and both relativistic and neutral reflection (\textsc{diskwind $\times$ relxill $+$ xillver}). We fit from 3--50~keV, and from 3--10~keV, to compare the relative constraining power. In each case, having established a best-fit, we then run an MCMC chain with 100 walkers for 200000 steps, after a burn in period of 100000 steps, to establish the parameter constraints.

\begin{figure*}
    \centering
    \includegraphics[height=0.4\linewidth]{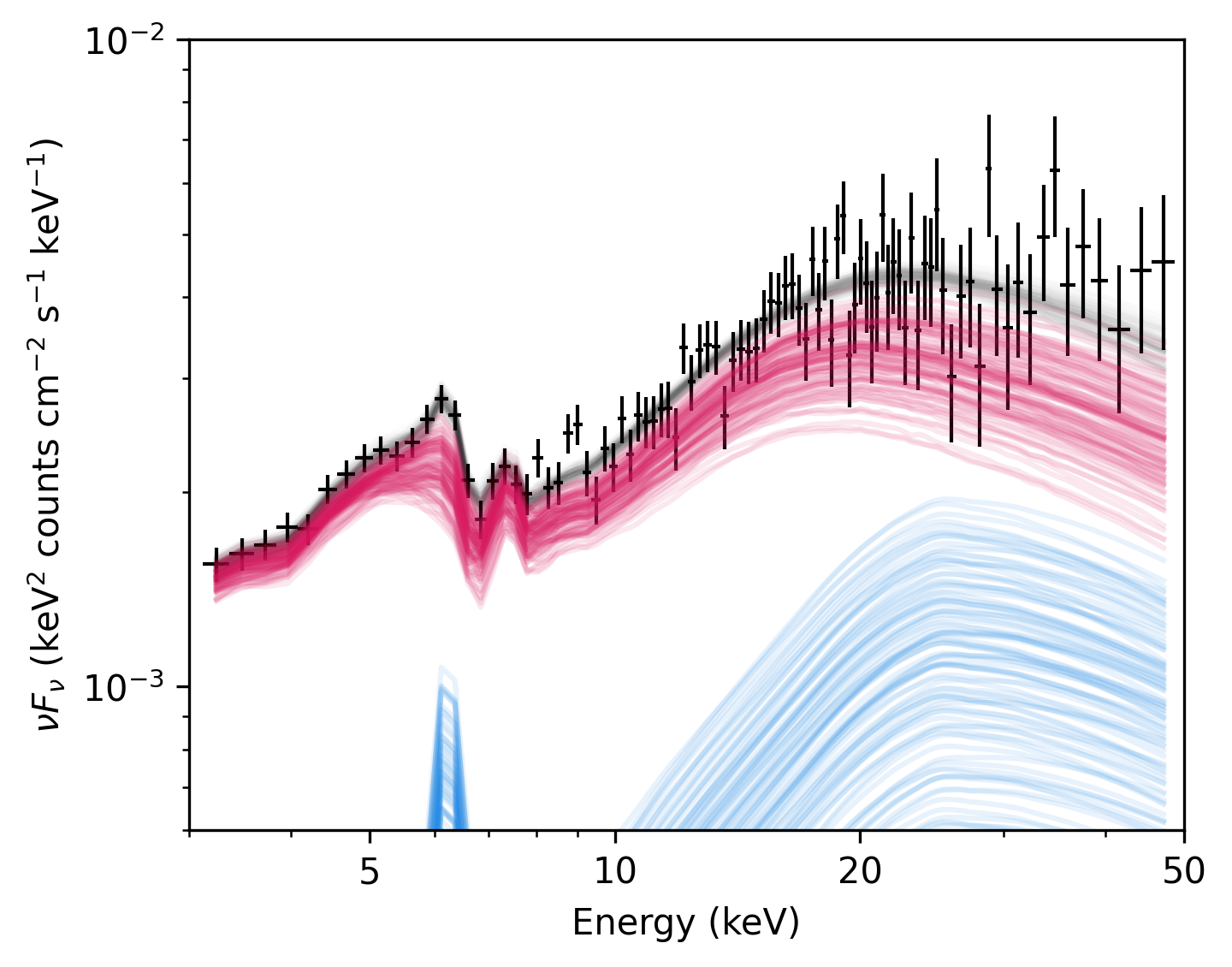}
    \includegraphics[height=0.4\linewidth]{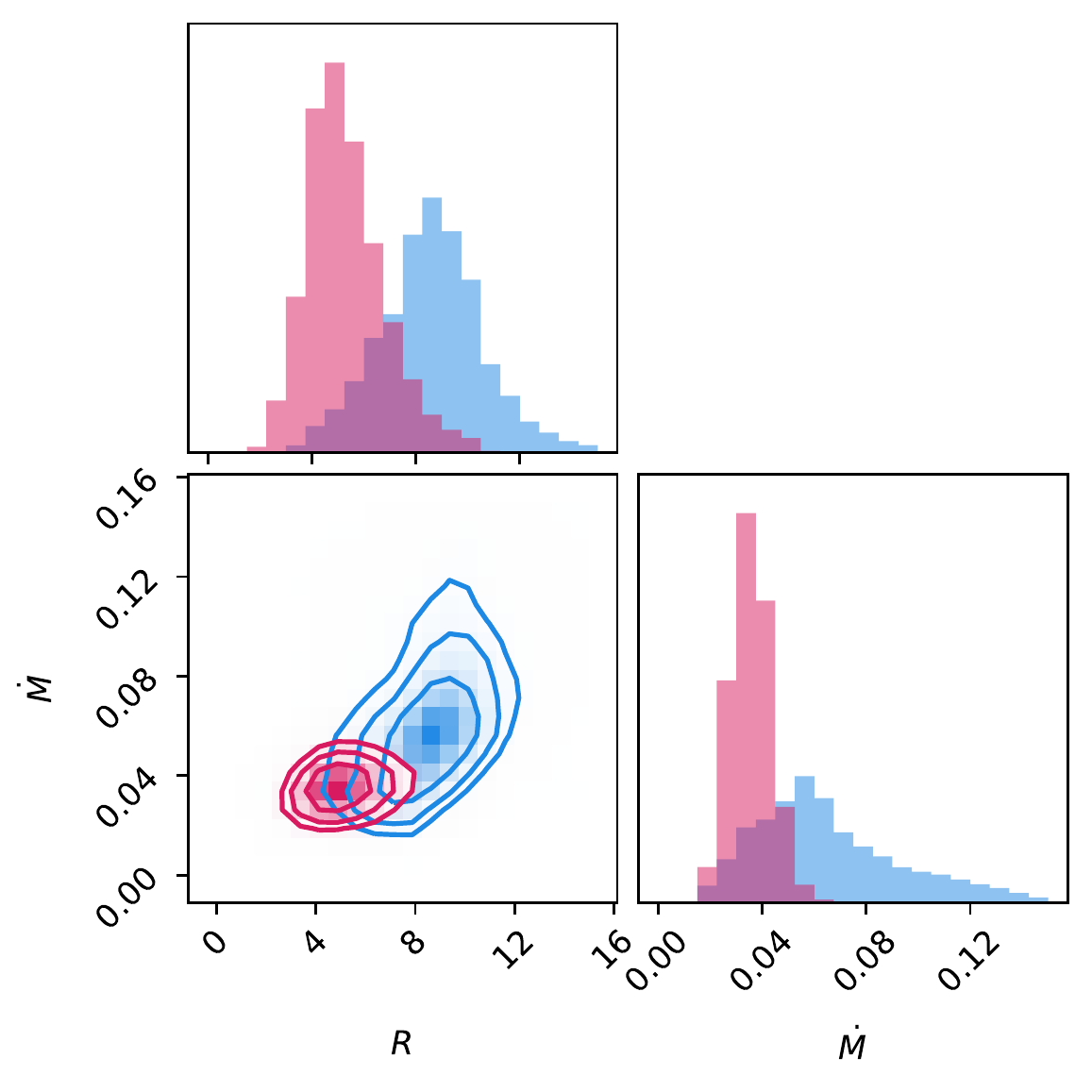}
    \caption{Left: Hybrid model fits to the \nustar\ spectrum of Mrk~335. 200 model lines drawn from MCMC chains are plotted, effectively showing the posterior distribution of model spectra. The red lines show the relativistic reflection component, modified by the disk wind, and the blue lines show the distant reflection. The relativistic reflection strength is strongly constrained by the need to fit the Compton hunp. Right: Posterios distributions of the reflection strength and mass outflow rate for the 3--50~keV band fit (red) and the 3--10~keV band fit (blue). Contours show the 1, 2 and 3 sigma limits. The parameter estimates derived from the full band fit are much more constrained than in the pure iron band fit.}
    \label{fig:mrk335}
\end{figure*}

\begin{figure}
    \centering
    \includegraphics[width=\linewidth]{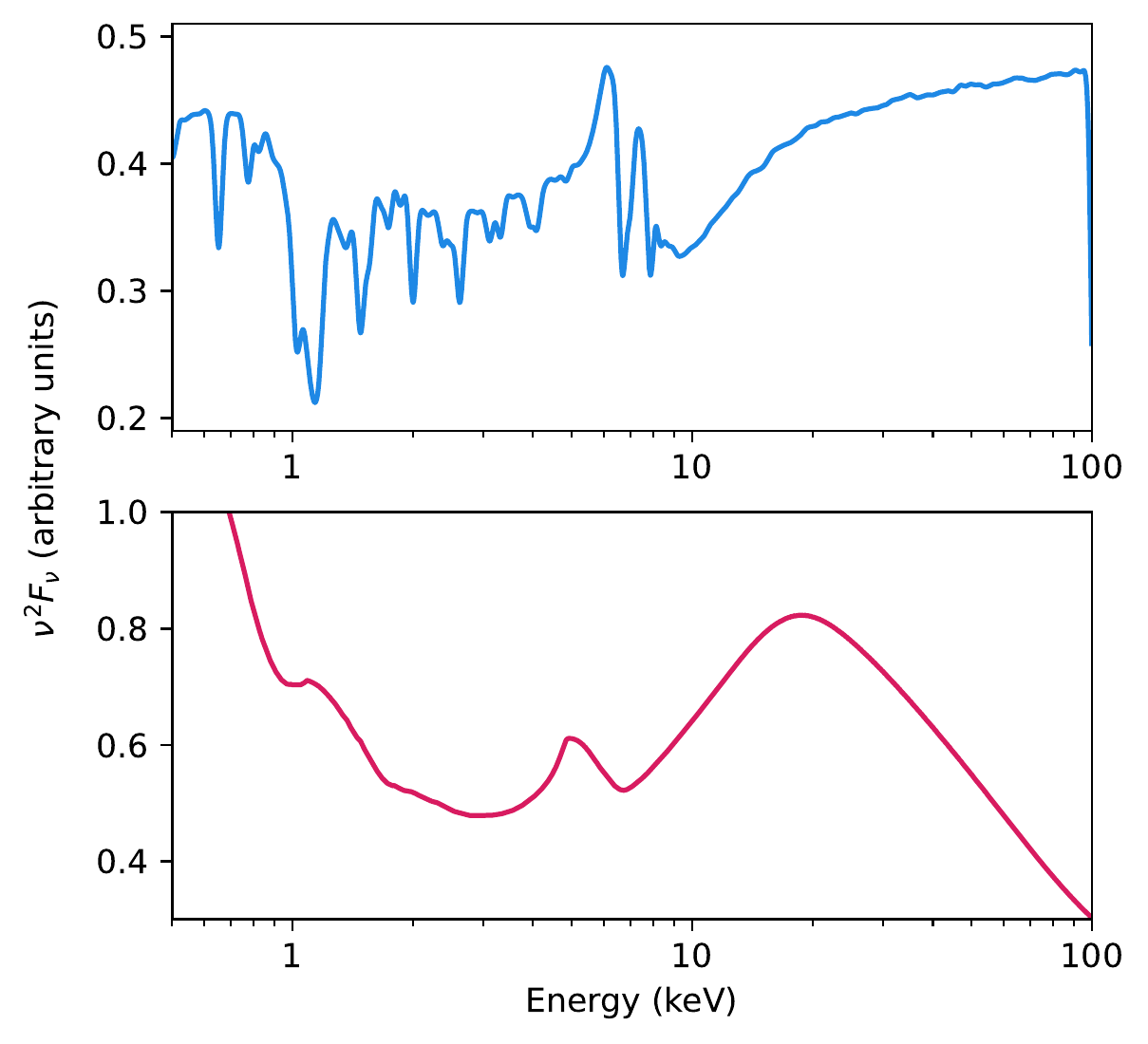}
    \caption{Breakdown of the model components from the best-fit to the Mrk~335 \nustar\ spectrum shown in Fig.~\ref{fig:mrk335}. The top panel shows the disk wind component, which contributes to the blue side of the iron line profile and adds some mildly blueshifted absorption lines. The bottom panel shows the reflection component, which produces most of the strongly redshifted part of the iron line and the Compton hump peaking at 20~keV. We apply a small Gaussian smoothing to the disk wind model for visual clarity.}
    \label{fig:mrk335_models}
\end{figure}

The full band data and 100 models form the MCMC chains are plotted in the left panel of Fig.~\ref{fig:mrk335}, and the posterior distributions of the reflection strength and mass outflow rates are plotted in the right panel. The full-band data offers a marked improvement over the 3--10~keV band, with much stronger constraints on the reflection strength and mass outflow rate. While this is a basic preliminary analysis, using a model that is not entirely self-consistent, the results are promising. In Fig.~\ref{fig:mrk335_models} we show a breakdown of the relativistic reflection and disk wind components, which demonstrates how the reflection component produces a strong Compton hump not seen in the disk wind model. This allows the relative strength of the two components to be constrained. We note that the strength of the iron line can vary independently of the Compton hump by changing the iron abundance, but this will affect other parts of the spectrum as well (e.g. the Fe~K edge, Fe~L features in the soft excess etc.).

The soft excess may also offer some additional constraining power, although this is debatable. Reflection models predict strong soft emission \citep[][]{Crummy06}, particularly with higher density disks \citep[e.g.][]{Garcia16, Jiang19}, which is not predicted by disk wind models. Additionally, for most X-ray detectors the soft excess contains the majority of the photon counts (and hence the majority of the signal), meaning that it can in principle offer great constraining power \citep[][]{Reynolds13Rev}. However, the nature of the soft excess in AGN is still hotly debated, in particular the relative contributions of warm Comptonisation and relativistic reflection \citep[e.g.][]{Middleton07, Jin13, Jin17, Garcia19, Petrucci20, Ballantyne20}. We note that in some cases there is evidence for strong Fe~L emission in the soft excess, matching the Fe~K line at high energies \citep[e.g.][]{Fabian09,Jiang18}, while in other cases there is a discrepancy between the parameters required to fit the soft excess and Fe~K line with reflection \citep[e.g.][]{Parker18_tons180}. It is possible that when the soft excess is better understood we will be able to use it to more reliably constrain the reflection contribution to the Fe~K line or measure spin directly, and \athena\ may help with this, but for now the soft excess is less useful.

\subsubsection{Multi-wavelength constraints}
Wind absorption signatures in the form of blue-shifted broad absorption lines (BALs) are present in rest-frame ultraviolet spectra of a significant fraction of luminous AGN \citep[e.g.][]{weymann1991,knigge2008,allen2011,rankine2020}, in addition to the X-ray features we have focused on. A disc wind or UFO will naturally be stratified in ionization state \citep{elvis_structure_2000,gallagher_stratified_2007,matthews_stratified_2020} due to absorption of the ionizing flux and density gradients within the flow -- as a result, the same outflow can, in principle, simultaneously imprint both absorption and emission features in the spectrum spanning a wide range of wavelengths. On the modelling side, this effect has been demonstrated explicitly using photoionization and radiative transfer modelling \citep{matthews_stratified_2020,mizumuto2021}. Observationally, a pertinent example is PDS 456, which shows UV absorption lines possibly associated with an extreme velocity outflow \citep{OBrien05,Hamann2018}. It is thus natural to ask whether an X-ray UFO will also produce wind signatures in other wavebands, and, more specifically, whether the presence or absence of these features can be used to discriminate between the wind and reflection scenarios.

To investigate multi-wavelength signatures, we simulated the UV spectrum using a different Monte Carlo radiative transfer code  optimised for the UV and optical portion of the spectrum (rather than the X-ray optimised code used by \citealt{Sim08,Sim10}).  The code is originally described by \cite{Long02}, and applied to modelling BAL quasar spectra by various authors \citep{Higginbottom13,matthews_testing_2016}. The fundamental techniques used in the two codes are very similar, and we use an identical wind geometry, adopting the best-fit wind parameters for PDS 456 from Matzeu et al. (in preparation); these parameters are $f_{\rm v}=1.25$, $L_{\rm X}/{\rm Edd}=0.005$ and $\dot{M}_{\rm out}/\dot{M}_{\rm Edd}=0.25$. We ran the simulation with and without radiation from a standard multi-temperature accretion disk with an innermost stable orbit of $6 r_g$; we discuss the pure power-law (without disk radiation) case here since this is the less ionized of the two. We found that the UV spectrum was featureless, consisting of a pure continuum without any atomic line features. The reason for the absence of features is that the wind is quite ionized, with $\log_{10} \xi > 4$ satisfied in the vast majority of the wind, such that there is not sufficient opacity in UV lines like C\textsc{iv}~1550\AA\ to form any absorption features. In fact, the dominant Carbon ion is C\textsc{vi} throughout the wind and the C\textsc{iv} ion fraction is $\lesssim 10^{-12}$ everywhere. Clearly, our modelling here is not an exhaustive effort, and it is entirely possible that relatively minor modifications to the wind parameters could result in UV wind features. For example, models with a wider range of launching angles could allow for shielding of material that launches from larger radii; this shielding effect would produce a more stratified ionization structure while preserving the inner ionized wind. Overall, our preliminary modelling suggests that while the presence of UV absorption features could feasibly be used to constrain wind models, the absence of such features cannot be used to rule out a wind contribution to the Fe K emission lines.

\subsubsection{Variability}

One key difference between wind and disk emission lies in the size of the emitting regions. For X-ray reflection, the bulk of the emission in the Fe~K line is coming from the innermost 10~$R_\mathrm{G}$, in particular the most strongly redshifted emission (below 5~keV) which originates deep in the black hole potential well. In a wind model, the emission region is much more extended, covering hundreds of gravitational radii. 

The logical implication of this is that emission from the wind should have lower amplitude, lower frequency variability than disk emission, as signals have to propagate over a much larger region. This should introduce a long delay, and average out high frequency variability. This is an oversimplification, however, as the light travel times and corresponding lags for emission from the approaching side of the disk can be small, allowing for rapid variability and short lags in the blueshifted emission \citep[e.g.][]{Mizumoto19}.

The receding part of the wind, on the other hand, should be much less variable, with a very long lag and a broad impulse response. In practise, that means that if a short, high frequency lag is observed in the redshifted emission, then it is almost certainly associated with disk reflection. We note that, regardless of wind parameters, the model discussed by \citet{Mizumoto19} does not predict any Fe~K lag below 6~keV. We note that several high profile detections of Fe~K lags show a delayed signal down to $\sim4$~keV \citep[e.g.][]{Kara13}, indicating that the redshifted emission must have a disk origin. Simultaneous lag and spectral modelling, as is made possible by models like \textsc{reltrans} \citep[][]{Ingram19}, may be an effective way to use the lag information to constrain the relative contributions of the wind and disk.

It should also be possibe to constrain the emission processes by frequency-resolving the lag. With wind reverberation, the line profile should move to higher energies at higher frequencies, and lower energies at lower frequencies, while the opposite should be true for reflection. This is obviously a technically challenging measurement to make, but we note that \citet{Zoghbi14} find tentative evidence of an Fe~K lag shifting redwards at higher frequencies in MCG-5-23-16.

These results argue for the presence of disk reflected emission in these sources, but do not necessarily rule out a contribution from winds, particularly at higher energies where the lags cannot be trivially distinguished. Fe~K reverberation is still not well understood, and Fe~K lags are not always present, even in sources that show clear reflection signatures \citep[][]{Kara14}. Ideally, we would be able to measure a lag signature at 5~keV, deduce that reflection must be present, and infer from that the amount of flux in the Fe~K line due to reflection, but this seems unlikely in the immediate future.

A more simplistic, but not necessarily worse, approach to using the variability information is to look at simpler diagnostics, like the $F_\mathrm{var}$ spectra \citep[e.g.][]{Vaughan03_variability}. Using the simple prediction that the redshifted emission in a wind model should be much less variable than the blueshifted emission, we may be able to derive a constraint on the amount of wind emission present by modelling $F_\mathrm{var}$ spectra. In our recent work on modelling these spectra \citep[e.g.][]{Parker20} we have shown that it is possible to derive the strength of the correlation between the continuum and reprocessed emission, however we did not examine this in detail or consider different correlations as a function of energy. This kind of study may be possible with current \xmm\ data, and will likely be trivial with the higher resolution and sensitivity of \athena .

\subsubsection{Ionisation changes}

In \citet{Parker17_nature} and \citet{Pinto18} we showed that the absorption lines from the UFO in IRAS~13224-3809 respond to the X-ray continuum, weakening as the flux rises in a manner consistent with ionisation of the gas. This has also been observed in PDS~456 \citep[][]{Parker18_pds456, Haerer21}, 1H~0707-495 \citep[][]{Parker21, Xu21}, and 30--60\% of a sample of variable AGN observed with \xmm\ \citep[][]{Igo20}, suggesting it is a general property of UFOs. If the ionisation interpretation of these observations is correct, then a similar effect may be visible in the emission from winds. This response would likely be more complex than that seen in absorption, but at high fluxes we might expect the wind emission lines to disappear altogether (with some delay for the redshifted side of the wind). The exact response would clearly depend on the geometry and ionisation structure of the wind, but this could easily be evaluated for specific wind geometries and tested observationally.

\subsubsection{Hybrid models}

We have discussed several methods for constraining the relative contributions of reflection and winds to the Fe~K emission profile. It is likely that none of these methods will guarantee that 100 per cent of the emission is from either process, so the logical approach to take is to use these constraints to provide priors for more sophisticated modelling, taking both processes into account self consistently. This should reduce the number of free parameters (and hence degeneracies), and will allow these systematic effects to be reflected in the estimated parameter uncertainties. 

The basic approach used in our simulations is a reasonable starting point: a reflection spectrum truncated at an outer radius corresponding to the wind launching radius, after which wind emission is produced. However, our set up assumes that both the disk and wind are illuminated by a pure powerlaw spectrum, which is not a realistic scenario. A self consistent model likely needs to include reflected emission in the input spectrum for the wind, since a significant fraction of the radiation hitting the wind is likely to have come from the disk first. The inverse is not necessarily true, if the corona is compact and the majority of the reflected emission comes from the innermost few gravitational radii, the amount of wind-scattered emission hitting the inner disk could easily be negligible compared to the coronal emission. 

\subsection{Caveats}
\label{sec:discuss_caveats}

In this study we have considered a single model set up, with one assumed geometry, and with model components that are not truly self-consistent. Considering all these factors is far beyond the scope of this paper, but means that the conclusions should be treated with caution. In particular, we suggest that any quantitative conclusions about the direction or magnitude of systematic biases should be treated as model dependent, and only the qualitative aspects (i.e. that potentially very large biases are introduced) should be treated as general results. Our knowledge of the geometry of both winds and the inner accretion flow is very limited, so we cannot be confident that the flat thin disk abruptly transitioning to a conical outflow implicitly assumed in our model is a valid approximation of the true geometry.

The disk wind model is calculated assuming illumination by a powerlaw spectrum, and this is assumed implicitly throughout when we use this model, including in our hybrid simulations. In a realistic disk wind scenario this is not the case - the wind will be illuminated both by the hot corona and by the reflected emission from the disk. Exactly what the effect of this will be is not yet known, and we intend to investigate this in future work.

To make this study practical certain simplifications had to be made to the fitting algorithm to ensure a reasonable run time for fitting 4000 spectra. In particular, we did not calculate the parameter errors on each fit, as this would have been computationally expensive and the errors typically reflect the scatter in the distribution of fit values. However, the \textsc{xspec} error calculation is frequently an effective way of getting fits away from a false minimum, so it is possible that a significant number of our fits are not located in the true minimum. To test this, we run the error calculations for 10 spectra from each set of fits, for a total of 40 spectra. For the \xmm\ reflection fits, we find no shift in the fit quality for the 7 well fit spectra, and a significant change in the fit statistic for the 3 poorly fit spectra. In one case, this is enough to shift it from being poorly fit to being well fit ($\chi^2_\nu$=2.3 to $\chi^2_\nu$=1.3), the others remain poorly fit despite the improvement. Fot the \xmm\ disk wind fits, two poorly fit spectra have a major change in the fit statistic, and for one of them it is enough to shift it into the well fit category, otherwise the changes are negligble. In both cases, none of the \athena\ spectra shift to a better fit after running the error calculations. This could indicate either that the true minimum is found more reliably with these data, or that false minima are significantly harder for the fit to escape. Overall, a conservative estimate is that $\sim10\%$ of the spectra that we label as poorly fit are actually well fit, with the fit stuck in a false minimum. In a manual fitting procedure, it is unlikely that these fits would have remained in their false minima, as users will almost always run the error calculations (and will generally try harder to make their model fit than an automated procedure).

We note that both the reflection and disk wind models have certain assumptions built in, such as the assumed wind geometry and the constant density and ionisation of the disk. These assumption effectively act as prior constraints on the models. This is particularly evident in the disk wind model, which has relatively few free parameters. This makes the parameter recovery more accurate in our tests, as the parameter space available is much smaller, but we stress that without genuine prior knowledge of the wind geometry this does not correspond to additional accuracy when fitting real data. Effectively, the scatter in the fit results is underestimated because the model is not free to vary fully.

We note that an alternative model for UFO absorption has been proposed, where the absorption lines originate in a layer of hot gas on the surface of the disk instead of a wind \citep[][]{Gallo11_ufos, Gallo13_ufos, Fabian2020_ufos}. In this case, the corresponding emission would be indistinguishable from conventional reflection, and would most likely result in a slight change in the emissivity profile without compromising parameter estimation. In fact, it is difficult to rule out this scenario precisely because of the degeneracy discussed here. If emission from large scale relativistic winds could be unambiguously detected then it would demonstrate that the absorption must be coming from the wind, rather than a disk surface layer.

\section{Conclusions}
\label{sec:conclusions}

We have investigated the effect of fitting pure reflection or disk wind models to a hybrid spectrum containing Fe~K emission from both processes. While the exact results we measure are likely strongly model and geometry dependent, some general conclusions can be drawn:
\begin{itemize}
    \item A large fraction of the simulated \xmm\ hybrid spectra were well fit in the iron~K band (3--10~keV) with the pure reflection and pure disk wind models, showing that it extremely difficult to distinguish the two models based on their emission lines alone.
    \item Higher quality \athena\ spectra do not break the degeneracy between the two models, as the broad emission lines are already resolved at CCD resolution.
    \item The spectra that were not well fit could likely be accommodated, in both reflection and disk wind cases, with relatively minor changes to the model, such as allowing for more complex absorption, a different wind opening angle, or distant reflection. In general, it is not possible to constrain the relative contributions of these two processes purely from Fe~K spectroscopy.
    \item If a single process model is assumed when both processes contribute to the spectrum, the parameters returned from that fit are likely to be strongly biased, and there is unlikely to be any evidence from spectroscopy alone that there is a problem with the model. The exact sense and amplitude of the biases introduced are presumably model dependent, but we have shown that in principle the systematic error can completely overwhelm the true signal.
    \item Without prior constraints on the relative contributions of wind and disk emission to the total emission profile, it is therefore impossible to reliably constrain key parameters from these models by fitting the Fe~K line.
\end{itemize}

We discuss various methods that can be used to mitigate this effect moving forward, such as broadband hybrid modelling, variability and multiwavelength constraints. The better we understand the system the stronger the priors we can put on the emission components, and the more reliable the parameter estimates returned will be. Overall, we are optimistic that this problem can be dealt with, but it will require a more sophisticated and comprehensive approach to spectral fitting than has been standard to date. We suggest that this should be a priority for the relevant communities in the build up to the launch of \xrism\ and \athena , so that we can maximise the scientific returns of these instruments.

\section*{Acknowledgements}
We thank the anonymous referee for their insightful and constructive comments, which have significantly improved the paper. J.H.M acknowledges a Herchel Smith Fellowship at Cambridge. J.J. acknowledges support from the Leverhulme Trust, the Isaac Newton Trust and St Edmund's College, University of Cambridge. MLP would like to thank Chris Reynolds and Andy Fabian for helpful discussions.

\section*{Data availability}
All \xmm\ and \nustar\ data used in this work are publicly available from the corresponding archives. Simulated spectra and automated analysis scripts are available on request to the authors. 




\bibliographystyle{mnras}
\bibliography{bibliography} 




\appendix

\section{Fit statistics and Athena}
\label{sec:chi2}

\begin{figure}
    \centering
    \includegraphics[width=0.8\linewidth]{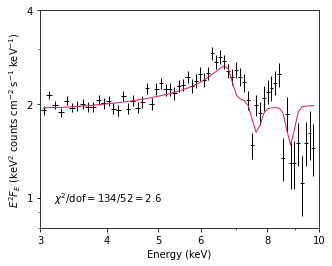}
    \includegraphics[width=0.8\linewidth]{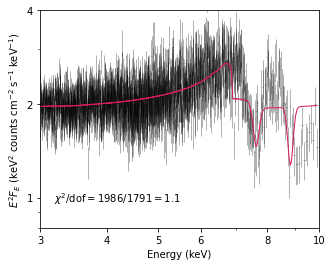}
    \includegraphics[width=0.8\linewidth]{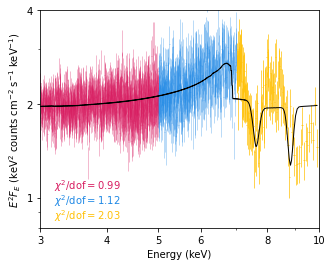}
    \caption{Top: Simulated \xmm\ spectrum with the hybrid model, fit with the reflection model. Middle: Simulated \athena\ spectrum with the same parameters, fit with the same model. Despite the much higher quality spectrum, $\chi^2_\nu$ is drastically lower. Bottom: Athena spectrum divided into three energy bands, with $\chi^2_\nu$ for each. This approach gives a better indication of where the fit is good and where it fails. }
    \label{fig:chi2}
\end{figure}

Throughout this work we use the reduced $\chi^2$ statistic to evaluate model fits to simulated data. This works reasonably well for the \xmm\ spectra, but an interesting problem arises when switching to the higher signal, higher resolution \athena\ spectra. Because of the drastically higher energy resolution, the \athena\ XIFU spectra have many more energy bins than \xmm , particularly at lower energies (<7~keV). Because this part of the spectrum is spectrally simple and relatively easy for both models to fit, this results in a much lower reduced $\chi^2$ value overall (this is not unique to $\chi^2$, the same effect occurs when using C-stat). We illustrate this for a single randomly selected spectrum in Fig.~\ref{fig:chi2}, where a spectrum with a very poor $\chi^2_\nu$ with \xmm\ data has a much lower $\chi^2_\nu$ with \athena\ data. A value of $\chi^2_\nu=2.6$ would generally be regarded as a poor fit, while $\chi^2_\nu=1.1$ would usually be treated as a reasonably good fit, and clearly this would be the wrong interpretation of these data. This effect is worsened by the steeper instrumental response of \athena\ relative to \xmm\ (the effective area of \athena\ decreases with energy more steeply than that of \xmm), which effectively means that lower energy data is weighted more highly by the fit statistic.

One way of avoiding this problem is to use a coarser binning for the \athena\ spectra. Binning to the same level as the \xmm\ data, for example, would essentially result in the same spectrum as the \xmm\ version but with much smaller error bars, and would give a very high value of $\chi^2_\nu$. However, this amounts to throwing away information to try and optimise the data for the fit statistic, which is clearly undesirable. Generally speaking, the data should not be binned more coarsely than the instrumental resolution, so long as an acceptable level of signal is achieved in each bin, otherwise information is lost.

It follows that if the data itself is not the problem, then the fit statistic is. The global reduced $\chi^2$, while very useful and simple to calculate, is not a perfect indicator of whether a model is providing a good description of data, particularly when the level of signal of the data varies strongly with energy. For our purposes, a better metric is to evaluate the $\chi^2$ in energy bands corresponding to specific spectral features of the model and data. We use three bands: 3--5~keV (the continuum), 5--7~keV (the iron emission line), and 7--10~keV (the absorption lines). This gives a much better indication of whether the fit is reasonable overall, while still being easy to automate without manual inspection. The bottom panel of Fig.~\ref{fig:chi2} illustrates this approach, which clearly identifies the poor fit to the absorption lines. This specific approach is unlikely to be generally applicable, but a similar technique based on checking local values of the reduced $\chi^2$ may be useful in many cases.

\section{Fit plots}
\label{sec:fitplots}


\begin{figure*}
    \centering
    \includegraphics[width=0.8\linewidth]{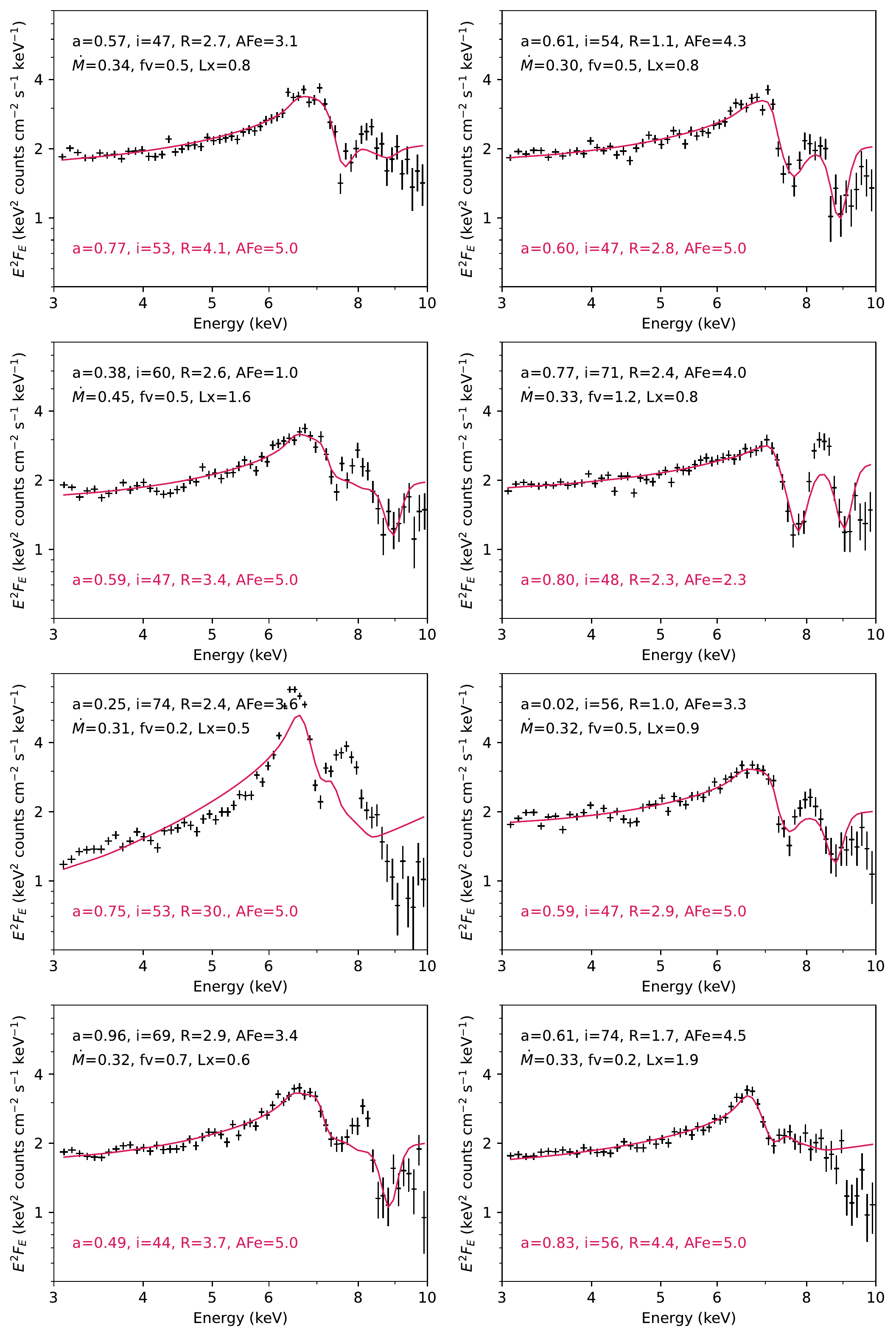}
    \caption{8 randomly selected \xmm\ spectra simulated with a hybrid reflection plus disk wind model, fit with a pure reflection model and two Gaussian absorption lines. These spectra are selected from the poorly-fit sample, with $\chi^2_\nu>2$. Some cases are due to a failure of the fitting algorithm, or complex absorption that our simple Gaussian model cannot describe well, but some are genuine cases of the reflection model being unable to fully describe the line profile. A common failure case is caused by a relatively narrow but double-peaked line profile. In practice, this could likely be better fit by including a distant reflection component in addition to the relativistic component.}
    \label{fig:ref_dirty_spectra}
\end{figure*}


\begin{figure*}
    \centering
    \includegraphics[width=0.8\linewidth]{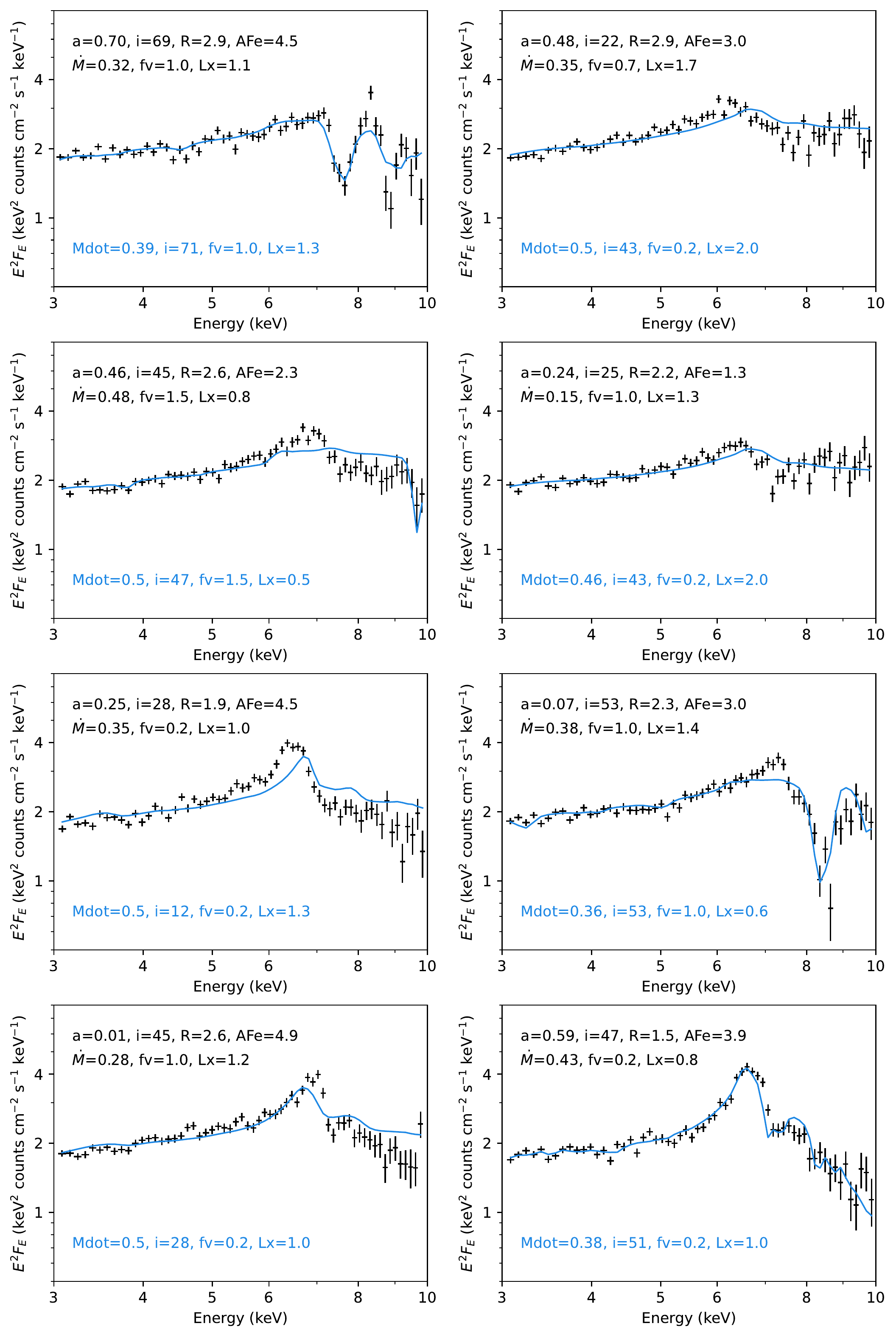}
    \caption{8 randomly selected \xmm\ spectra simulated with a hybrid reflection plus disk wind model, fit just the disk wind model. These spectra are selected from the poorly-fit sample, with $\chi^2_\nu>2$. Most of these cases occur when the line profile is sharply peaked but with a net red- or blue-shift.}
    \label{fig:ref_dirty_spectra}
\end{figure*}



\begin{figure*}
    \centering
    \includegraphics[width=0.8\linewidth]{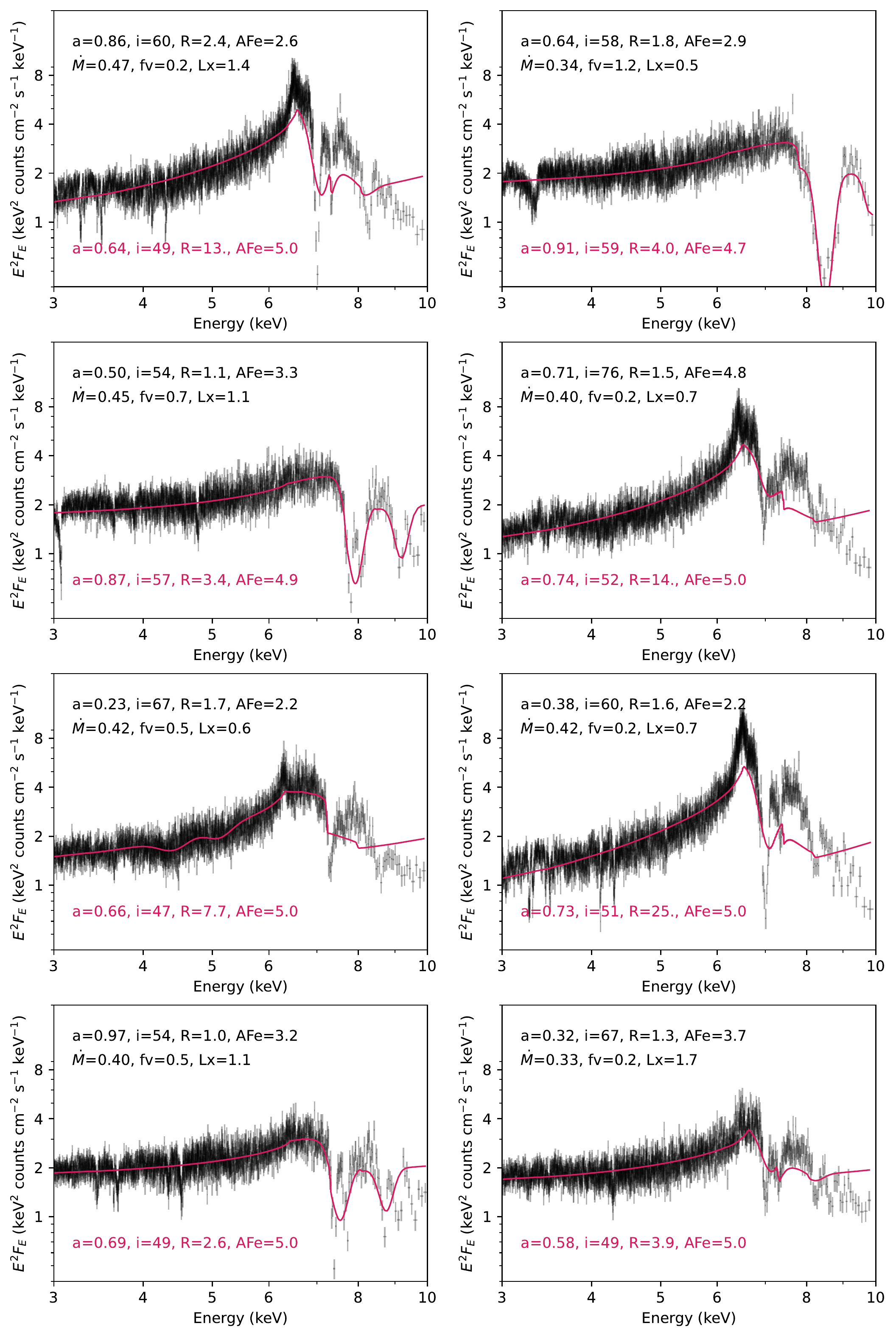}
    \caption{8 randomly selected \athena\ XIFU spectra simulated with a hybrid reflection plus disk wind model, fit with a pure reflection model and two Gaussian absorption lines. These spectra are selected from the poorly-fit sample, with $\chi^2_\nu>1.4$. Most of these cases occur when the simplistic absorption model fails, rather than when the model cannot adequately describe the emission profile.}
    \label{fig:ref_dirty_spectra}
\end{figure*}



\begin{figure*}
    \centering
    \includegraphics[width=0.8\linewidth]{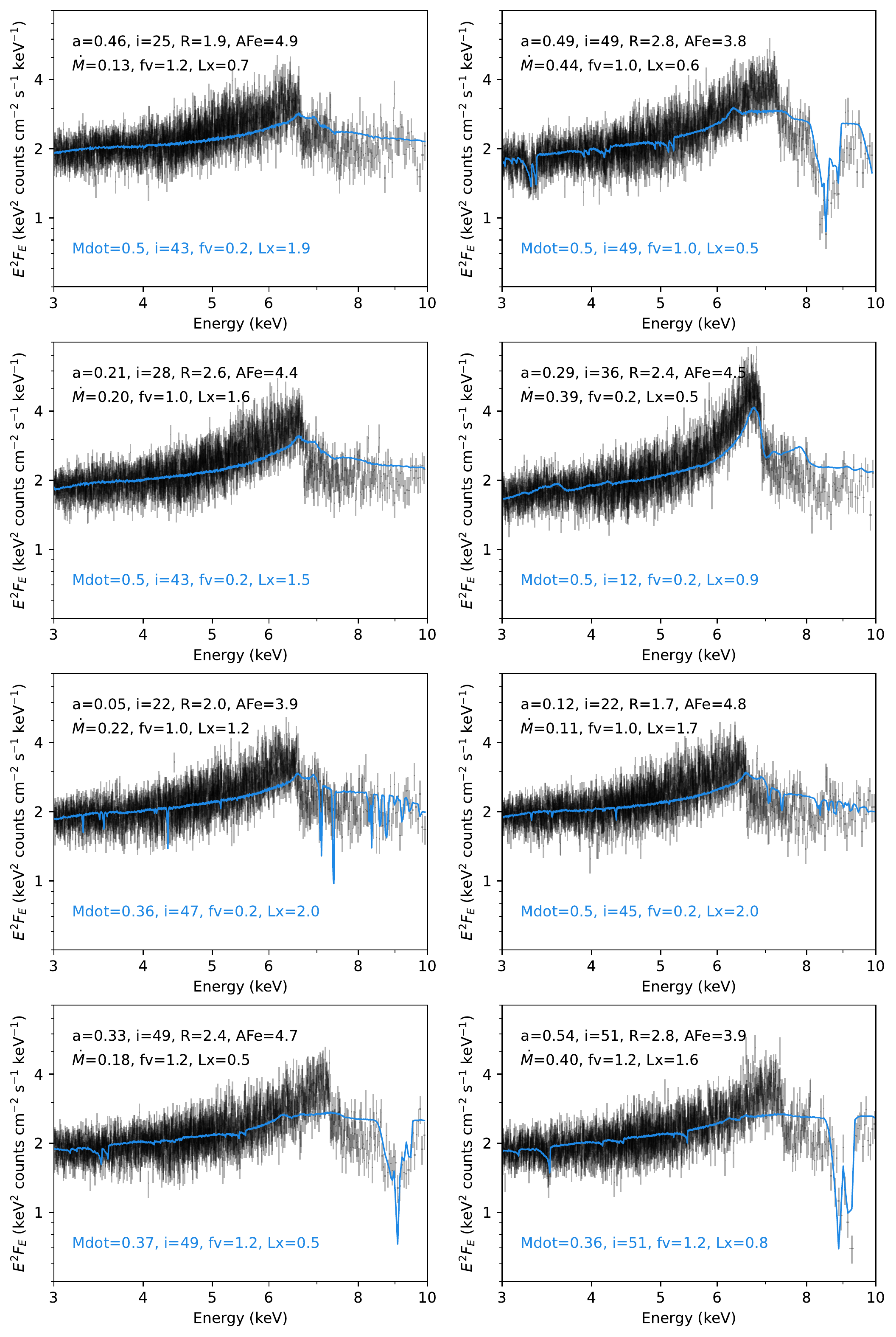}
    \caption{8 randomly selected spectra XIFU simulated with a hybrid reflection plus disk wind model, fit with just the disk wind model. These spectra are selected from the poorly-fit sample, with $\chi^2_\nu>1.4$. These cases are very rare (only 12 of the 1000 simulated spectra), and mostly seem to be due to a failure of the fitting algorithm.}
    \label{fig:ref_dirty_spectra}
\end{figure*}


\bsp	
\label{lastpage}
\end{document}